\documentclass[prl,twocolumn,showpacs,preprintnumbers,amsmath,amssymb, superscriptaddress]{revtex4-1}

\usepackage[makeroom]{cancel}

\usepackage{amsmath}    
\usepackage{amssymb}
\usepackage{graphicx}   
\usepackage{verbatim}   
\usepackage{color}      
\usepackage{hyperref}   
\usepackage[normalem]{ulem}
\usepackage{natbib}
\usepackage{fixmath}
\usepackage{enumitem}

\hypersetup{colorlinks,linkcolor=blue,urlcolor=blue,citecolor=blue}
\newcommand{\beq}{\begin{equation}}
\newcommand{\eeq}{\end{equation}}
\newcommand{\bea}{\begin{eqnarray}}
\newcommand{\eea}{\end{eqnarray}}

\newcommand{\be}{\begin{equation}}
\newcommand{\ee}{\end{equation}}
\newcommand{\Nedge}{N_{\rm edge}}

\definecolor{darkgreen}{rgb}{0,0.5,0}
\definecolor{orange}{rgb}{1,0.5,0}
\definecolor{grey}{rgb}{.6,.6,.6}
\newcommand{\rc}[1]{\textcolor{black}{#1}}

\newcommand{\bS}{\mathbf S}

\newcommand{\ba}{\mathbf a}
\newcommand{\bC}{\mathbf C}
\newcommand{\bH}{\mathbf H}

\newcommand{\bx}{\mathbf x}
\newcommand{\br}{\mathbf r}
\newcommand{\brp}{{\mathbf r}'}

\begin{document}
\title{Topologically protected, correlated end spin formation in carbon nanotubes }
        
\author{C\u at\u alin Pa\c scu Moca}
\affiliation{MTA-BME Quantum Dynamics and Correlations Research Group, 
Institute of Physics, Budapest University of Technology and Economics, 
Budafoki \'ut 8., H-1111 Budapest, Hungary}
\affiliation{Department of Physics, University of Oradea, 410087, Oradea, Romania}
\author{Wataru Izumida}
\affiliation{Department of Physics, Tohoku University, Sendai 980-8578, Japan}
\author{Bal\' azs D\' ora}
\affiliation{Department of Theoretical Physics and MTA-BME Lend\"ulet Topology and Correlation Research Group,
Budapest University of Technology and Economics, 1521 Budapest, Hungary}
\author{\" Ors Legeza}
\affiliation{Strongly Correlated Systems Lend\" ulet Research Group, Institute for Solid State Physics and Optics,
  {Wigner Research Centre} for Physics, P.O. Box 49, H-1525 Budapest, Hungary}
\author{{J\'anos K. Asb\'oth}}
\affiliation{BME-MTA Exotic Quantum Phases 'Lend\"ulet' Research Group, Institute of Physics, Budapest University of Technology and Economics, 
Budafoki \'ut 8., H-1111 Budapest, Hungary}
\author{Gergely Zar\'and}
\affiliation{MTA-BME Quantum Dynamics and Correlations Research Group, 
Institute of Physics, Budapest University of Technology and Economics, 
Budafoki \'ut 8., H-1111 Budapest, Hungary}
\affiliation{BME-MTA Exotic Quantum Phases 'Lend\"ulet' Research Group, Institute of Physics, Budapest University of Technology and Economics, 
Budafoki \'ut 8., H-1111 Budapest, Hungary}
\date{\today}
\begin{abstract}
For most chiralities, semiconducting nanotubes display topologically protected end states of multiple degeneracies. 
We demonstrate  using density matrix renormalization group
based quantum chemistry tools that 
 the presence of Coulomb interactions induces the formation of  robust end spins. These are the close 
analogues of ferromagnetic edge states emerging in graphene nanoribbons. The interaction between 
the two ends is sensitive to the length of the nanotube, its dielectric constant, as well as the size of the end spins: 
for $S=1/2$ end spins their interaction is antiferromagnetic, while for $S>1/2$ it changes from antiferromagnetic to 
ferromagnetic  as the nanotube length increases. The interaction between end spins can be controlled  by changing the dielectric constant of the environment, 
 thereby providing a possible platform for two-spin quantum manipulations.
\end{abstract}
\maketitle

\paragraph{Introduction --} 
Topological insulators represent unique states of matter, and besides their theoretical appeal, 
they hold    promise for revolutionizing  quantum computation, spintronics and thermal electrics~\cite{hasankane,zhangrmp,xu2017}.
While their insulating bulk does not differ significantly from that of a simple band insulator, their topological character is manifested by the appearance of emergent
surface and edge states, frequently exhibiting unusual physical properties. 
Probably the best known incarnation of a topological state is the edge 
state in the Su-Schrieffer-Heeger (SSH) model~\cite{SSH}, describing the dimerization of
polyacetylene
. In this case,  the dimerized phase is a topological band insulator, and correspondingly, 
at the edges of the
polyacetylene chain or at topological defects separating different dimerized phases, mid-gap 
bound states and corresponding local spin excitations  emerge~\cite{Li.2014, Meier.2016}. 

Although nanotubes have continuously been in the focus of extremely intense research for more than two decades by now
\cite{Charlier.2007,Ayala.2010, Kouwenhoven.2015,Donarini2019,Hills2019,Khivrich2019, Shapir.2019,Grifoni.2019, Graf2017,Hata.2018}, surprisingly, it has been discovered only recently  that most  insulating carbon nanotubes
also belong to the class of topological systems. As a consequence, they should 
possess mid-gap states~\cite{Ilani,Wataru.2016, Wataru.2018}, quite similar to those found in the  SSH model. Quite astonishingly, 
as we discuss below, the number and character of these mid-gap states is exclusively determined by 
the chirality of the nanotube, and in most nanotubes, \emph{several} end states are predicted to appear at each 
end of the tube. However, in a neutral and non-interacting nanotube, all these states would be almost degenerate, 
and therefore they are expected to be most sensitive to interaction effects.

\begin{figure}[b!]
	\includegraphics[width=0.95\columnwidth]{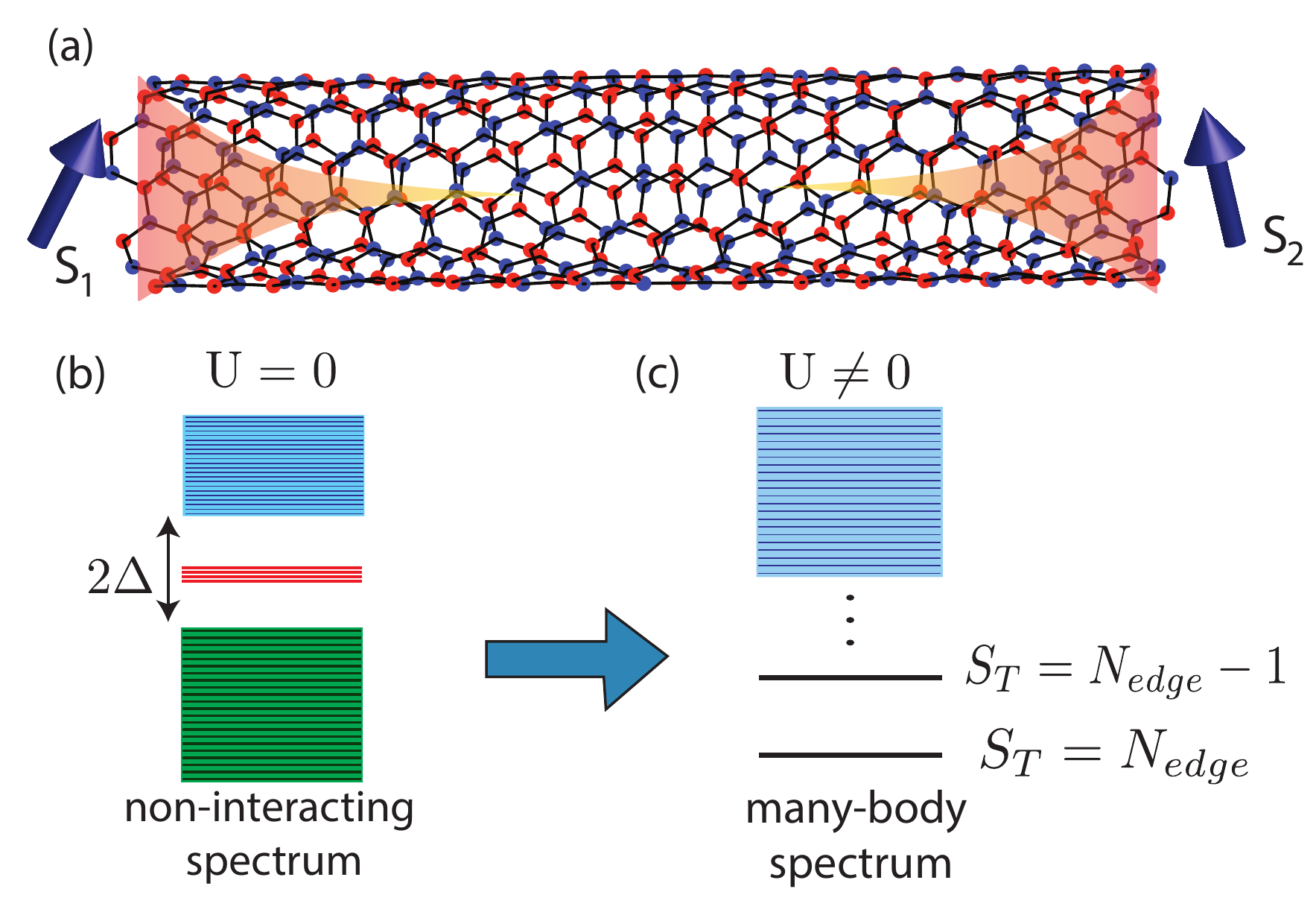}
	\caption{ (a)  Topologically protected spins are formed at both edges of most semiconducting nanotubes.
	(b)  Band structure of a semiconducting nanotube in the absence of interactions. Topological end states (red lines) appear in the  gap.  
	(c)  Many-body spectrum at finite interaction. For ferromagnetic end spin coupling,  the ground state has a total spin $S_T$,  equal to the number of edge states
	$N_{\rm edge}$. Spin excitations appear at low energies due to coupling between end spins. }
	\label{fig:edge_spins}
\end{figure}

In this work, we focus our attention to these  interaction effects, and demonstrate that 
--- in the presence of interactions --- these topologically protected end states behave in many ways as spontaneously 
formed quantum dots.   In particular, 
interactions lead to spin formation   and tend to align spins ferromagnetically
at each end of the nanotube~\cite{Higuchi_2004, Hod2008, Lin, Mananes}, thereby producing end spins of size 
\begin{equation}
S_1=S_2 = \frac{N_{\rm edge} } 2,
\label{eq:endspin}
\end{equation}
with $N_{\rm edge}$ denoting the total number of topologically protected mid-gap states at each end (see Fig.~\ref{fig:edge_spins}.(a)).  
Depending on chirality,  $N_{\rm edge}$ can be quite large  for many nanotubes, implying the 
appearance of surprisingly large end spins, paralleling in many ways ferromagnetic edge states observed 
in graphene nanoribbons~\cite{Kunstmann,Wakabayashi_2010, Dutta2012, Yazyev, Magda2014, Tao2011}. 
The two end spins then couple to each other via an exchange interaction which, in the absence of spin-orbit coupling,
 takes on a simple form: 
\begin{equation}
H_{\rm exch} = {1\over 2}J_{\rm eff}\, { \bS}_1\, {\bS}_2\;.
\label{eq:excha}
\end{equation}
The sign and strength of the exchange interaction here turns out  to depend sensitively on the length of the 
nanotube as well as on its chirality and the dielectric constant of its environment. 
\rc{Spin-orbit coupling is not expected to influence spin formation, it will, however, lead to some degree of exchange anisotropy, and also induce local spin anisotropies. As a result, the SU(2) degenerate spin multiplets are expected to split, and for long nanotubes the end $S>{1\over 2}$ spins will behave rather as coupled Ising spins.}

\paragraph{Hamiltonian. --} 
In this work, we  use a tight binding approach to describe interacting nanotubes~\cite{Reich}, and express the Hamiltonian as
\begin{align}
H  =  -\sum_s \sum_{\br, \brp} t(\br-\brp)\,c^\dagger_s(\br) c_s(\brp)    
\nonumber
\\
 +{1\over 2} \sum_{\br, \brp } V(\br-\brp) :n(\br):: n(\brp):.
 \label{eq:Hamiltonian}
\end{align}
Here  $c^\dagger_s(\br)$
creates an electron with spin $s$ at the $p_z$ orbital of a carbon atom at 
a position $\br$.  The hopping matrix elements $ t(\br-\brp)$ describe hopping between nearest neighbour and next nearest neighbour orbitals. They incorporate curvature effects~\cite{KouwenhovenRMP}, and also can be generalized to include spin-orbit effects 
neglected here~\cite{Kuemmeth2008}.  
The second term in Eq.~\eqref{eq:Hamiltonian} accounts for the long-ranged Coulomb interaction
between local charge fluctuations on the nanotube
\begin{equation}
 V ( \br) 
  = \frac{e^2}{\epsilon} \frac{1}{ \sqrt{ \br^2 + {\alpha}^2 } },
\end{equation}
with $U_0 = 11.3\, \rm{eV}$and $ \alpha  \approx 0.127 \, \text{nm}/\epsilon$ a short distance cut-off, and $\epsilon$ the dielectric constant~\cite{Perebeinos-PRL-2004}.
Densities in Eq.~\eqref{eq:Hamiltonian} appear in a normal ordered form, $ :n(\br):\,\equiv \sum_s
(c^\dagger_s(\br) c_s(\br)-1/2)$, thereby measuring deviations from half filling. 
In the following, we shall determine and analyze the many-body ground state and excitation spectrum of this Hamiltonian.  

\paragraph{Non-interacting nanotubes and  topological end states.---} 
Nanotubes are classified by their chirality, $\chi = (n,m)$, i.e. the lattice vector  ${\bf  C} = n {\bf a}_1 + m {\bf a}_2 $,
along  which a graphene sheet needs to be rolled up to form the nanotube. 
In this work, we focus on semiconducting nanotubes with $(n-m) \,\text{mod}\, 3 = \pm1 $.

For topological considerations, it is most useful to consider a perfect and infinite nanotube, and 
use a so-called helical construction~\cite{White, Jishi}.  Similar to graphene, the nanotube  
possesses two sublattices, $A$ and $B$. In  the helical construction, 
one introduces a helical  vector $\bf H$ within the graphene sheet, 
and lines up all atoms of the nanotube along just $d$ spirals  along the direction $\bf H$, 
with $d$ defined as  the greatest common divisor  of $n$ and $m$ (see supplemental material 
\footnote{See Supplementary material for details on the effective 1D lattice model
  and DMRG calculations which includes Refs.~\cite{Reich, White, Jishi,KouwenhovenRMP,Perebeinos-PRL-2004}} for details.
 
 \begin{figure}[t!]
	\includegraphics[width=1\columnwidth]{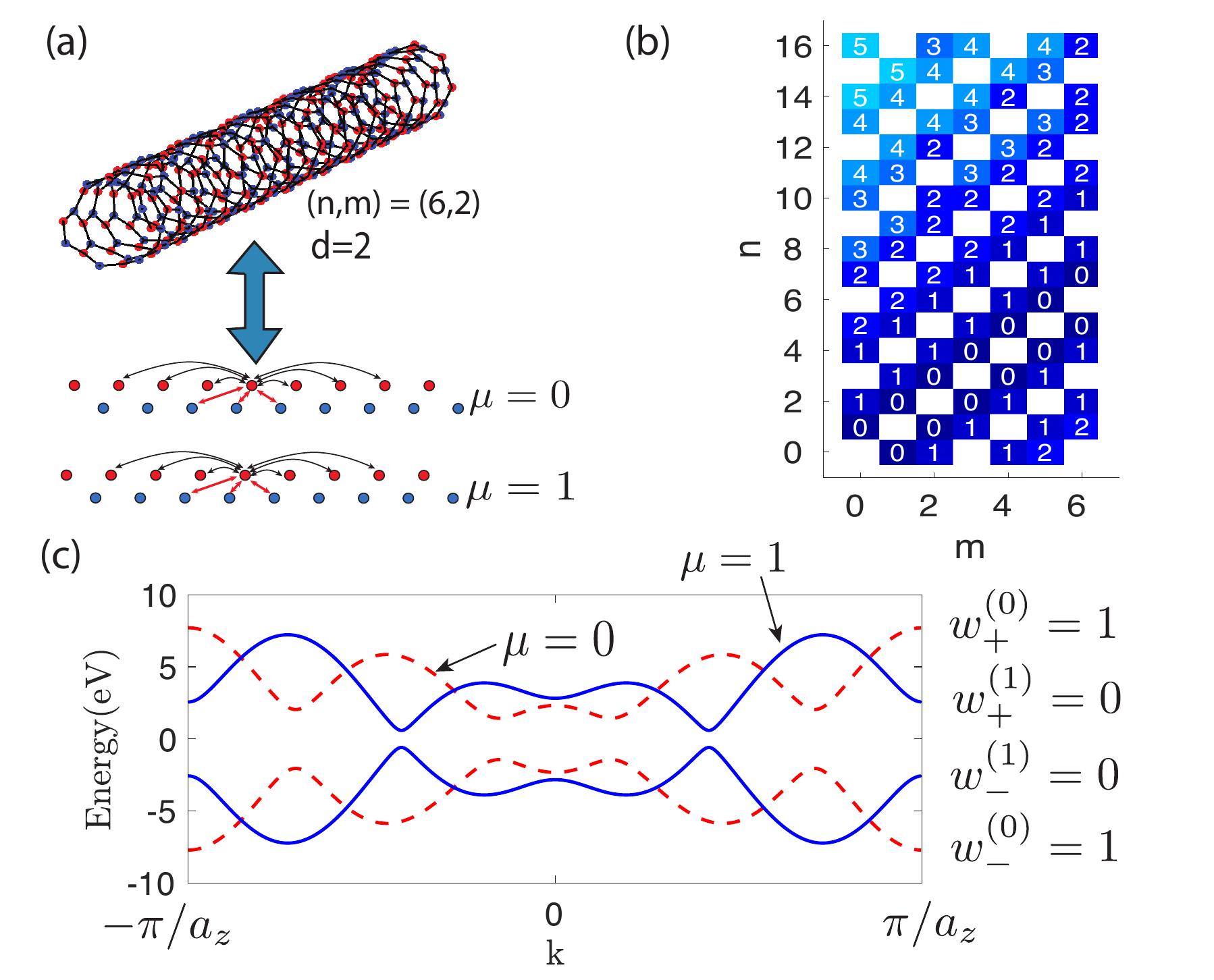}
	\caption{ 
	(a) Mapping of an infinite  carbon nanotube to an effective 1D ladder-like lattice model with 
	$d$ decoupled chains, for  a chirality $\chi=(6,2)$ and $d= 2$.
	 Arrows  indicate hoppings between carbon atoms. (b) The number of edge states, 
	 $N_{\rm edge}$, as a function of the chirality $\chi=(n,m)$. 
	 (c) Band structure and the corresponding winding numbers  for a  $(6,2)$ nanotube. 
	  }
	  
	\label{fig:non-interacting}
\end{figure}
 
 Clearly, an infinite nanotube  possesses a  discrete $d$-fold rotational symmetry around the axis of the tube, $C_d$, 
 and a 'gliding' (helical) translational symmetry along the chain, as generated by the helical vector $\bf H$.
Correspondingly, single particle (but also many-body) states can be
labeled by their "angular momentum" $\mu=0,\dots,d-1$ and a
quasimomentum $k$ along the chain, and are organized into $2d$ bands,
$\epsilon_{\pm}^{(\mu)}(k)$, the band index $\pm$ originating from the
sublattice structure of the nanotube, and referring to bonding
(valence) and anti-bonding (conduction) bands.

Within the tight binding scheme used here, these bands are associated with $d$ independent 
one dimensional chains, each giving rise to one conduction and one valence band, and 
describing the motion    of electrons with a given 'angular momentum' $\mu$
(see Fig.~\ref{fig:non-interacting}(c)). Interestingly, each of these bands possesses 
a topological winding number~\cite{Wataru.2016}  $w^{(\mu)}$. 
Non-zero winding numbers imply the presence of topologically protected end 
states~\cite{Hatsugai,teo2010topological,Wataru.2018}. 
Remarkably, we can express
the total number of end states at each end of a semiconducting tube
 in a closed form, just in terms of the nanotube's chirality,
\begin{multline}
N_{\rm edge}=2\Big \lfloor {n-m \over 3d}\Big \rfloor + 3\Big \lfloor {d+1 \over 3}\Big \rfloor +2\,\Theta(d) -\\
\Big ( \Big \lfloor {d+1 \over 3}\Big \rfloor  +\Theta(d)  \Big)\Theta( {n-m \over d} )\;,
 \label{eq:N_edge}
\end{multline}
where  $\Theta(x) = (x+1) \,{\rm mod}\, 3 -1$
 is a modified modulo function taking values $0$ and $\pm1$, 
 and  $\lfloor\dots \rfloor$ denotes the floor function. In Fig.~\ref{fig:non-interacting}.(b), we 
 display $N_{\rm edge}$ as a function of the chirality of the nanotubes. White squares indicate metallic tubes, 
 while colored ones refer to semiconducting tubes.  Clearly, most of the tubes are semiconducting, and the vast majority  
 of semiconducting tubes  possess topological end states, typically several ones. For zig-zag tubes with chirality $(n,0)$, e.g.,
the number of end states increases linearly with the circumference of the tube, $N_{\rm edge}^{\text{zig-zag}} \approx n/3$.  

Remarkably, as our tight binding calculations also demonstrate, these
end states are rather robust and not very sensitive to the form of the
ending of the nanotube as long as it terminates in a minimal edge,
i.e., with the minimal number of missing atoms and dangling bonds per
period~\cite{akhmerov2008boundary}. This is due to the fact that end
states extend over many lattice sites, both along the circumference
and along the width of the nanotube, hence defects at the end of the
nanotube that break sublattice symmetry or the $C_d$ symmetry mix end
states only slightly. Note that end states are not robust against
sublattice-selectively removing or adding some atoms at an end: this
removes or creates end states, and alters the size and interaction of the end spins, accordingly.
However, such modifications of the nanotube break the minimal edge condition, and are thus energetically
unfavorable~\cite{akhmerov2008boundary}.

\paragraph{Interacting nanotubes.---}
To perform numerical calculations, we first construct a finite nanotube, and diagonalize the noninteracting 
part of the Hamiltonian Eq.~\eqref{eq:Hamiltonian} to find its eigenstates $\phi_\alpha(\br)$
and the corresponding eigenenergies, $\epsilon_\alpha$, and express the interaction term
within this basis. Normal ordering needs to be treated with special care in this process.
To treat nanotubes of reasonable length, $L\approx 40\, \text{nm}$, we restrict the many-body calculations 
to just about a hundred active states  from the valence and conduction bands with energies 
$|\epsilon_\alpha| < \Lambda \approx 5 \Delta$, with $\Delta$ the band gap of the non-interacting infinite nanotube. 
Then we apply  a density matrix renormalization group (DMRG) based approach adopted to 
Hamiltonians with arbitrary long-ranged two-body interactions~\cite{WhiteDMRG, Schollwock, Ors} to determine the 
ground state and low lying excitations  of the  nanotube. In this procedure, we use $U(1)\times U(1)$ symmetries, i.e. we 
fix the excess charge $ Q$ on the nanotube and the $z$ component of the total spin, $S_T^z$.
In practice, the computational basis is further optimized using fermionic mode transformation\cite{Krumnow}.
\begin{figure}[b!]
	\includegraphics[width=0.9\columnwidth]{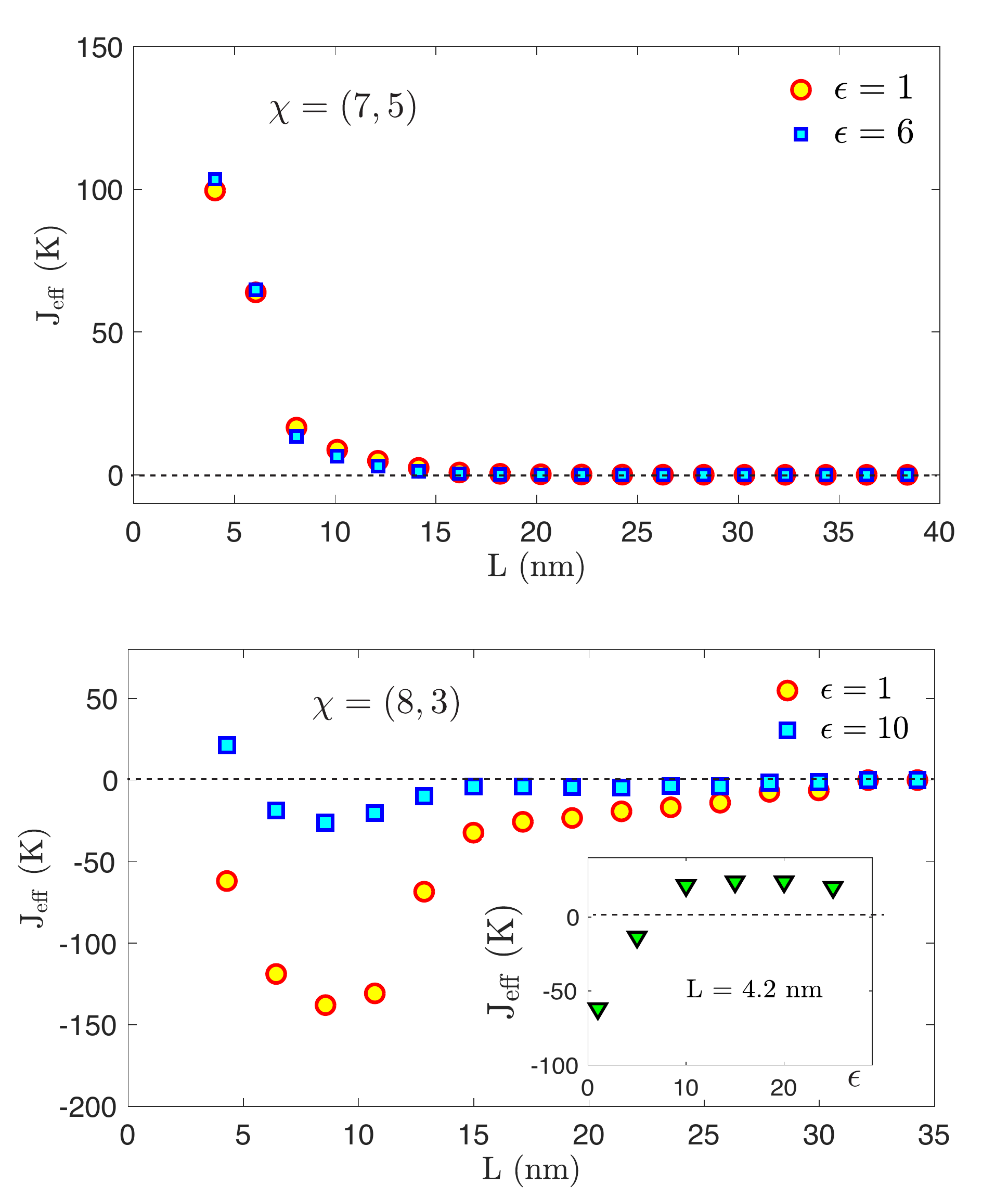}
	\caption{ Effective exchange interaction $J_{\rm eff}$ between the localized spins  at the two ends of the nanotube as function of its length. 
	When $N_{edge} $=1, $J_{eff} $ is always positive indicating an antiferromagnetic exchange, while for $N_{edge}\ge 2$  an antiferromagnetic
	 to ferromagnetic transition occurs. As the inset shows, for appropriate nanotube length, the sign of the 
	 interaction can be changed by changing the dielectric constant of the environment.}
	\label{fig:J_eff}
\end{figure}

As sketched in Fig.~\ref{fig:edge_spins},  end spins manifest  in the form 
of low energy sub-gap excitations, which can be described by the effective Hamiltonian, 
Eq.~\eqref{eq:excha}. The many-body spectra observed   reveal consistently the formation of 
end spins with $S_{1,2}=\Nedge/2$, coupled to each other. In the absence of spin-orbit coupling, 
this interaction is $SU(2)$ symmetrical, and  the many-body spectrum 
consists of multiplets with total spin $S_T=0,\dots, N_{\rm edge}$. 

The alignment and size of the electron spins at the ends of the nanotubes can be easily understood. 
In a topological nanotube, $2\Nedge$ spin degenerate states are split from the conduction and val\textcolor{red}{e}nce bands, 
and form the mid-gap states, and are therefore populated by $2\Nedge$ electrons in a neutral (half-filled)  tube.
End states are thus half-filled in a neutral nanotube.
The spatial extension of these localized end states states is roughly $\xi_0\sim \hbar c/\Delta \sim R$, with 
$c$ the Fermi velocity and $R$ the radius of the nanotube. Electrons confined on these states interact therefore strongly with each other, and moving one electron from one end of the tube to the other would cost an energy 
$\sim E_C\sim e^2/(\epsilon \xi_0) \sim e^2/(\epsilon R) $. Therefore, to minimize their Coulomb energy,  $\Nedge$ electrons go  to each end of the tube. Moreover, since all these single particle levels are degenerate, and wave functions on one end overlap with each other, electrons at one end follow Hund's rule, and align their spins to minimize their interaction, thereby yielding a composite spin, $S_{1,2}=\Nedge /2$, Eq.~\eqref{eq:endspin}.

We have analyzed the excitation spectra of dozens of nanotubes, and  verified Eq.~\eqref{eq:endspin}
numerically in the presence of Coulomb interaction for all nanotubes listed in Fig.~\ref{fig:non-interacting}.b. 
In these simulations, we have observed end spins as large as $S_{1,2}=5/2$, and  corresponding ground state spins 
  as large as $S_T=S_1+S_2=5$. According  to  Eqs.~\eqref{eq:endspin}
and \eqref{eq:N_edge},  for appropriate  chiralities and larger 
  nanotube radii, the total emergent spin can largely exceed these values. 
The ground state spin of the nanotube is determined by the 
exchange coupling $J_{\rm eff}$ between the end spins. Being generated by 
tunneling between the topological end states, this coupling is  expected  to fall off 
exponentially with the length of the nanotube.  The  coupling $J_{\rm eff}$
 can be readily extracted from the spin excitation spectrum, and is displayed for 
two particular nanotubes as a function the nanotube length $L$ in Fig.~\ref{fig:J_eff}.  On  top,
 we show the results for  a $(7,5)$ nanotube with $N_{\rm edge}=1$, and corresponding spin 
 $S=1/2$'s at the edges. The coupling is antiferromagnetic, and therefore $S_T=0$ in this case, irrespective 
 of the length of the nanotube. As expected, the coupling  $J_{\rm eff}$ decays exponentially with $L$, reflecting the exponentially  localized nature of the end states. 
 
 A completely different behavior is observed, however,  for an $(8,3)$ nanotube with $\Nedge=2$, 
 as displayed on the bottom of Fig.~\ref{fig:J_eff}.
  Here we observe an antiferromagnetic coupling in very short nanotubes
 with $L\lesssim 5\,\text{nm}$, while in longer tubes the interaction becomes ferromagnetic 
 and  decays exponentially, as expected. 
 
 The behavior shown in Figs.~\ref{fig:J_eff} appears to be  generic. We have studied 
 a great number of  nanotubes with different chiralities, and in all nanotubes with $\Nedge=1$ we 
 find an antiferromagnetic coupling, while all nanotubes with $\Nedge\ge 2$ exhibit an exchange interaction 
 that changes from antiferromagnetic to ferromagnetic with increasing nanotube length. 
 As demonstrated 
 in the lower panel, the  precise location of the sign change is sensitive to 
 the dielectric constant, $\epsilon$, and by appropriate engineering of $\epsilon$, one can even completely 
 decouple  the two end spins. This mechanism provides a tool to perform quantum manipulations 
 with the end spins.

\paragraph{Charging the end states.---}
As discussed above, a topological  nanotube behaves to a large extent as a self-organized 
double quantum dot system. Whether one can charge these topological quantum dots
 or not and observe the  end states in a direct spectroscopic (tunneling) experiment, depends 
largely on screening, i.e., the value of $\epsilon$. Placing an additional electron to the topological states 
costs a Coulomb energy of the order of $E_C\sim e^2/(\epsilon R)$, while adding a delocalized particle to the 
valence band needs an energy $\Delta \sim \hbar c/ R$. Therefore, for each chirality, there is a critical 
value $\epsilon_C\sim e^2/(\hbar c)$ of the dielectric constant. For dielectric constants larger than $\epsilon_C$ (strong screening), electrons and holes added to a neutral nanotube localize at  the end and the 
topological quantum dots can be charged, while for smaller dielectric constants (weak screening) they must 
go directly to the conduction or valence band, and delocalize along the nanotube. 

\begin{figure}[t!]
	\includegraphics[width=0.9\columnwidth]{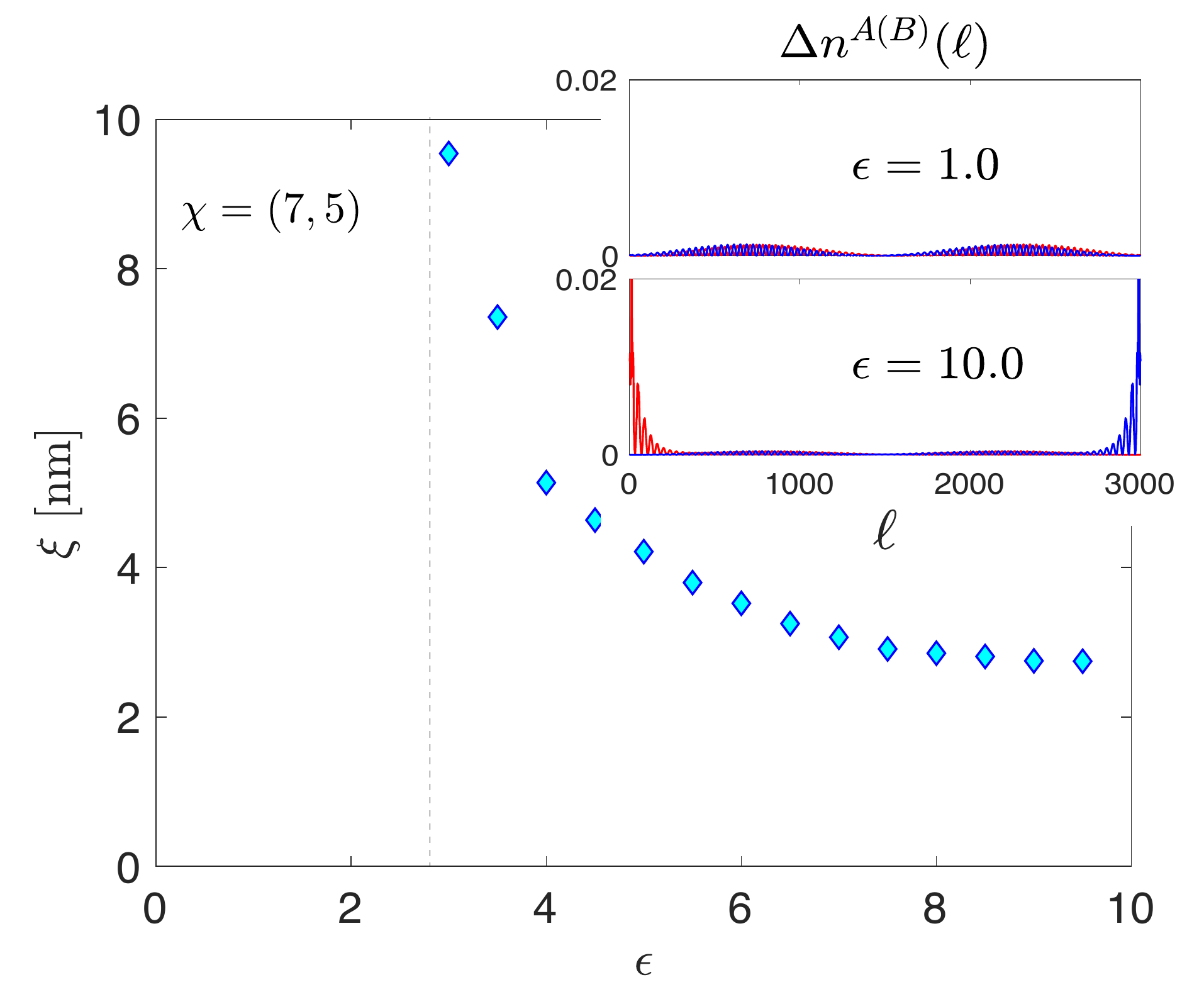}
	\caption{ Extension of the wave function of the additional spectrum, as a function of $\epsilon$
	in  a  $(7,5)$ nanotube of length {$L=30\,\text{nm}$}, with $\Nedge=1$. 
	For $\epsilon< \epsilon_C\approx 2.5$, the added charge delocalizes along the nanotube, while for 
	$\epsilon> \epsilon_C$ the charge is added to the topological quantum dots (it is delocalized between them 
	in the ground state). As the coloring indicates, the two end states live on different sublattices.  
	The localization length of the added electron diverges as $\epsilon$ approaches $\epsilon_C$.}
	\label{fig:xi}
\end{figure}

According to our calculations, this  transition happens at around $\epsilon \approx 3$, as is displayed in Fig.~\ref{fig:xi}. 
The inset of Fig.~\ref{fig:xi}
shows the spatial location of an  electron added to the nanotube in terms of 
the position $\ell$ along the helix.  Clearly, the added particle is localized on sublattice $A$ at one end, 
while it localizes on sublattice $B$ at the other end (in close similarity with the SSH model). 
As shown in the main panel, the localization length of the  added particle, $\xi$ is strongly 
influenced by Coulomb interactions, and diverges as one approaches the 
critical value of $\epsilon$. This localization length should not be confused with that of the end spins, which 
remains of the order of $R$. Close to $\epsilon\gtrsim \epsilon_C$, the delocalized charges can create a glue 
between the end spins. 

\paragraph{Closing observations and conclusions.---}
As we demonstrated in this work, most carbon nanotubes are topological, and 
 all topological nanotubes possess interaction induced  end spins, residing 
 at the edges of the tube, and localized within a distance $\sim \xi_0\sim R$. 
Being protected by topology,  these naturally formed end spins are robust, 
 are typically larger than spin $S=1/2$, and couple to each other 
exponentially weakly in longer nanotubes (longer than a few nanometers). 
Their presence may provide a natural explanation for the  intrinsic spin formation 
 observed longtime ago in  encapsulated nanotubes (pea pods) ~\cite{Ferenc}, and
simple model calculations support that   an exponentially 
 weak ferromagnetic exchange quite naturally explains the super-Curie
 behavior reported earlier~\cite{nafradi}.
 
 The large end spins demonstrated  here are the nanotube  analogues of  ferromagnetic   edge states 
appearing in graphene nanoribbons~\cite{Kunstmann, Dutta2012, Yazyev, Magda2014, Tao2011, Ors2016}. Indeed, selecting 
\emph{any} topologically non-trivial chirality $(p,q)$ with $p$ and $q$ being relative primes,  we can think of nanoribbons of width $W$ as  nanotubes with chirality $(n,m) \equiv (r\,p,r\,q)$ and length $L=W$, with  $r$ taken to infinity. 
In this limit, the length of the nanotube remains finite while its radius $R$ is taken to infinity, thereby yielding 
nanoribbons  closed into a cylinder. 
In this limit, the rotational symmetry $C_{d\to\infty}$ yields a proliferation of topological end states, 
 thus forming a dispersionless band that is subject to Stoner ferromagnetism.
The sign change of $J_{\rm eff}$
observed has also its counterpart in nanoribbons:   in close analogy
 with the sign change  of $J_{\rm eff}$ observed here, the coupling between  ferromagnetic  edge states is  observed to change sign, too,  from being antiferromagnetic to ferromagnetic as a function of   $W$~\cite{ Magda2014}.
 
Topological nanotubes spontaneously form double dot devices, which may provide a platform for 
quantum computation.  As we demonstrated, local 
 probes such as scanning tunneling microscopy (STM) can be used to observe these  'topological quantum dots' , 
 however, to charge them, the effective dielectric constants must be increased over some 
 critical  value. Therefore, rather then using suspended nanotubes, nanotubes layed 
 over some tunable dielectrics would be the most promising candidates for a direct experimental 
 observation by tunneling spectroscopy.  
 \rc{Another way to detect these protected end states may be via local optical spectroscopy. Excitonic states, i.e. bound sub-gap electron-hole excitations have been observed by two-photon spectroscopy in bulk nanotubes \cite{ Wang.2005, Maultzsch.2005}. Charging the end states, and binding a charge carrier of opposite sign to it should create similar excitonic edge states. These edge excitons should have a binding energy clearly distinct from that of the bulk excitons, and may be detected by optically probing the edge of the nanotube in the infrared. Direct edge state $\to$ valence or conduction band} excitations should also be possibly detected well below the optical gap, $E_g=2\Delta$.

\emph{Acknowledgments.--} We thank Ferenc Simon,  Christoph Strunk,   Levente Tapaszt\'o, and 
P. Nemes-Incze for  insightful  discussions.  
This work has been supported by the National Research, Development and Innovation Office (NKFIH) through 
 Grant Nos. K119442 and K120569, 
through  the Hungarian Quantum Technology National Excellence Program, project no. 2017-1.2.1-NKP-2017- 00001, 
by the BME-Nanotechnology FIKP grant  (BME FIKP-NAT), by the OTKA grant number FK 132146
and by  the  Romanian National Authority for Scientific Research and Innovation, UEFISCDI, under 
project no. PN-III-P4-ID-PCE-2016-0032. W.I. acknowledges  support from 
KAKENHI Grants Nos. JP15K05118, JP15KK0147, JP18H04282.
\"O.L. also acknowledges financial support from the Alexander von Humboldt foundation.

Ö.L. acknowledges  support from the Center for Scalable and Predictive Methods 
for Excitation and Correlated Phenomena (SPEC), funded by the Computational Chemical 
Sciences Program of the U.S.  Department of Energy (DOE).

\bibliography{references}

\begin{thebibliography}{48}%
\makeatletter
\providecommand \@ifxundefined [1]{%
 \@ifx{#1\undefined}
}%
\providecommand \@ifnum [1]{%
 \ifnum #1\expandafter \@firstoftwo
 \else \expandafter \@secondoftwo
 \fi
}%
\providecommand \@ifx [1]{%
 \ifx #1\expandafter \@firstoftwo
 \else \expandafter \@secondoftwo
 \fi
}%
\providecommand \natexlab [1]{#1}%
\providecommand \enquote  [1]{``#1''}%
\providecommand \bibnamefont  [1]{#1}%
\providecommand \bibfnamefont [1]{#1}%
\providecommand \citenamefont [1]{#1}%
\providecommand \href@noop [0]{\@secondoftwo}%
\providecommand \href [0]{\begingroup \@sanitize@url \@href}%
\providecommand \@href[1]{\@@startlink{#1}\@@href}%
\providecommand \@@href[1]{\endgroup#1\@@endlink}%
\providecommand \@sanitize@url [0]{\catcode `\\12\catcode `\$12\catcode
  `\&12\catcode `\#12\catcode `\^12\catcode `\_12\catcode `\%12\relax}%
\providecommand \@@startlink[1]{}%
\providecommand \@@endlink[0]{}%
\providecommand \url  [0]{\begingroup\@sanitize@url \@url }%
\providecommand \@url [1]{\endgroup\@href {#1}{\urlprefix }}%
\providecommand \urlprefix  [0]{URL }%
\providecommand \Eprint [0]{\href }%
\providecommand \doibase [0]{http://dx.doi.org/}%
\providecommand \selectlanguage [0]{\@gobble}%
\providecommand \bibinfo  [0]{\@secondoftwo}%
\providecommand \bibfield  [0]{\@secondoftwo}%
\providecommand \translation [1]{[#1]}%
\providecommand \BibitemOpen [0]{}%
\providecommand \bibitemStop [0]{}%
\providecommand \bibitemNoStop [0]{.\EOS\space}%
\providecommand \EOS [0]{\spacefactor3000\relax}%
\providecommand \BibitemShut  [1]{\csname bibitem#1\endcsname}%
\let\auto@bib@innerbib\@empty
\bibitem [{\citenamefont {Hasan}\ and\ \citenamefont {Kane}(2010)}]{hasankane}%
  \BibitemOpen
  \bibfield  {author} {\bibinfo {author} {\bibfnamefont {M.~Z.}\ \bibnamefont
  {Hasan}}\ and\ \bibinfo {author} {\bibfnamefont {C.~L.}\ \bibnamefont
  {Kane}},\ }\href {\doibase 10.1103/RevModPhys.82.3045} {\bibfield  {journal}
  {\bibinfo  {journal} {Rev. Mod. Phys.}\ }\textbf {\bibinfo {volume} {82}},\
  \bibinfo {pages} {3045} (\bibinfo {year} {2010})}\BibitemShut {NoStop}%
\bibitem [{\citenamefont {Qi}\ and\ \citenamefont {Zhang}(2011)}]{zhangrmp}%
  \BibitemOpen
  \bibfield  {author} {\bibinfo {author} {\bibfnamefont {X.-L.}\ \bibnamefont
  {Qi}}\ and\ \bibinfo {author} {\bibfnamefont {S.-C.}\ \bibnamefont {Zhang}},\
  }\href {\doibase 10.1103/RevModPhys.83.1057} {\bibfield  {journal} {\bibinfo
  {journal} {Rev. Mod. Phys.}\ }\textbf {\bibinfo {volume} {83}},\ \bibinfo
  {pages} {1057} (\bibinfo {year} {2011})}\BibitemShut {NoStop}%
\bibitem [{\citenamefont {Xu}\ \emph {et~al.}(2017)\citenamefont {Xu},
  \citenamefont {Xu},\ and\ \citenamefont {Zhu}}]{xu2017}%
  \BibitemOpen
  \bibfield  {author} {\bibinfo {author} {\bibfnamefont {N.}~\bibnamefont
  {Xu}}, \bibinfo {author} {\bibfnamefont {Y.}~\bibnamefont {Xu}}, \ and\
  \bibinfo {author} {\bibfnamefont {J.}~\bibnamefont {Zhu}},\ }\href@noop {}
  {\bibfield  {journal} {\bibinfo  {journal} {npj Quantum Materials}\ }\textbf
  {\bibinfo {volume} {2}},\ \bibinfo {pages} {51} (\bibinfo {year}
  {2017})}\BibitemShut {NoStop}%
\bibitem [{\citenamefont {Su}\ \emph {et~al.}(1979)\citenamefont {Su},
  \citenamefont {Schrieffer},\ and\ \citenamefont {Heeger}}]{SSH}%
  \BibitemOpen
  \bibfield  {author} {\bibinfo {author} {\bibfnamefont {W.~P.}\ \bibnamefont
  {Su}}, \bibinfo {author} {\bibfnamefont {J.~R.}\ \bibnamefont {Schrieffer}},
  \ and\ \bibinfo {author} {\bibfnamefont {A.~J.}\ \bibnamefont {Heeger}},\
  }\href {\doibase 10.1103/PhysRevLett.42.1698} {\bibfield  {journal} {\bibinfo
   {journal} {Phys. Rev. Lett.}\ }\textbf {\bibinfo {volume} {42}},\ \bibinfo
  {pages} {1698} (\bibinfo {year} {1979})}\BibitemShut {NoStop}%
\bibitem [{\citenamefont {Li}\ \emph {et~al.}(2014)\citenamefont {Li},
  \citenamefont {Xu},\ and\ \citenamefont {Chen}}]{Li.2014}%
  \BibitemOpen
  \bibfield  {author} {\bibinfo {author} {\bibfnamefont {L.}~\bibnamefont
  {Li}}, \bibinfo {author} {\bibfnamefont {Z.}~\bibnamefont {Xu}}, \ and\
  \bibinfo {author} {\bibfnamefont {S.}~\bibnamefont {Chen}},\ }\href {\doibase
  10.1103/PhysRevB.89.085111} {\bibfield  {journal} {\bibinfo  {journal} {Phys.
  Rev. B}\ }\textbf {\bibinfo {volume} {89}},\ \bibinfo {pages} {085111}
  (\bibinfo {year} {2014})}\BibitemShut {NoStop}%
\bibitem [{\citenamefont {Meier}\ \emph {et~al.}(2016)\citenamefont {Meier},
  \citenamefont {An},\ and\ \citenamefont {Gadway}}]{Meier.2016}%
  \BibitemOpen
  \bibfield  {author} {\bibinfo {author} {\bibfnamefont {E.~J.}\ \bibnamefont
  {Meier}}, \bibinfo {author} {\bibfnamefont {F.~A.}\ \bibnamefont {An}}, \
  and\ \bibinfo {author} {\bibfnamefont {B.}~\bibnamefont {Gadway}},\ }\href
  {https://doi.org/10.1038/ncomms13986} {\bibfield  {journal} {\bibinfo
  {journal} {Nature Communications}\ }\textbf {\bibinfo {volume} {7}},\
  \bibinfo {pages} {13986} (\bibinfo {year} {2016})}\BibitemShut {NoStop}%
\bibitem [{\citenamefont {Charlier}\ \emph {et~al.}(2007)\citenamefont
  {Charlier}, \citenamefont {Blase},\ and\ \citenamefont
  {Roche}}]{Charlier.2007}%
  \BibitemOpen
  \bibfield  {author} {\bibinfo {author} {\bibfnamefont {J.-C.}\ \bibnamefont
  {Charlier}}, \bibinfo {author} {\bibfnamefont {X.}~\bibnamefont {Blase}}, \
  and\ \bibinfo {author} {\bibfnamefont {S.}~\bibnamefont {Roche}},\ }\href
  {\doibase 10.1103/RevModPhys.79.677} {\bibfield  {journal} {\bibinfo
  {journal} {Rev. Mod. Phys.}\ }\textbf {\bibinfo {volume} {79}},\ \bibinfo
  {pages} {677} (\bibinfo {year} {2007})}\BibitemShut {NoStop}%
\bibitem [{\citenamefont {Ayala}\ \emph {et~al.}(2010)\citenamefont {Ayala},
  \citenamefont {Arenal}, \citenamefont {Loiseau}, \citenamefont {Rubio},\ and\
  \citenamefont {Pichler}}]{Ayala.2010}%
  \BibitemOpen
  \bibfield  {author} {\bibinfo {author} {\bibfnamefont {P.}~\bibnamefont
  {Ayala}}, \bibinfo {author} {\bibfnamefont {R.}~\bibnamefont {Arenal}},
  \bibinfo {author} {\bibfnamefont {A.}~\bibnamefont {Loiseau}}, \bibinfo
  {author} {\bibfnamefont {A.}~\bibnamefont {Rubio}}, \ and\ \bibinfo {author}
  {\bibfnamefont {T.}~\bibnamefont {Pichler}},\ }\href {\doibase
  10.1103/RevModPhys.82.1843} {\bibfield  {journal} {\bibinfo  {journal} {Rev.
  Mod. Phys.}\ }\textbf {\bibinfo {volume} {82}},\ \bibinfo {pages} {1843}
  (\bibinfo {year} {2010})}\BibitemShut {NoStop}%
\bibitem [{\citenamefont {Laird}\ \emph
  {et~al.}(2015{\natexlab{a}})\citenamefont {Laird}, \citenamefont {Kuemmeth},
  \citenamefont {Steele}, \citenamefont {Grove-Rasmussen}, \citenamefont
  {Nyg\aa{}rd}, \citenamefont {Flensberg},\ and\ \citenamefont
  {Kouwenhoven}}]{Kouwenhoven.2015}%
  \BibitemOpen
  \bibfield  {author} {\bibinfo {author} {\bibfnamefont {E.~A.}\ \bibnamefont
  {Laird}}, \bibinfo {author} {\bibfnamefont {F.}~\bibnamefont {Kuemmeth}},
  \bibinfo {author} {\bibfnamefont {G.~A.}\ \bibnamefont {Steele}}, \bibinfo
  {author} {\bibfnamefont {K.}~\bibnamefont {Grove-Rasmussen}}, \bibinfo
  {author} {\bibfnamefont {J.}~\bibnamefont {Nyg\aa{}rd}}, \bibinfo {author}
  {\bibfnamefont {K.}~\bibnamefont {Flensberg}}, \ and\ \bibinfo {author}
  {\bibfnamefont {L.~P.}\ \bibnamefont {Kouwenhoven}},\ }\href {\doibase
  10.1103/RevModPhys.87.703} {\bibfield  {journal} {\bibinfo  {journal} {Rev.
  Mod. Phys.}\ }\textbf {\bibinfo {volume} {87}},\ \bibinfo {pages} {703}
  (\bibinfo {year} {2015}{\natexlab{a}})}\BibitemShut {NoStop}%
\bibitem [{\citenamefont {Donarini}\ \emph {et~al.}(2019)\citenamefont
  {Donarini}, \citenamefont {Niklas}, \citenamefont {Schafberger},
  \citenamefont {Paradiso}, \citenamefont {Strunk},\ and\ \citenamefont
  {Grifoni}}]{Donarini2019}%
  \BibitemOpen
  \bibfield  {author} {\bibinfo {author} {\bibfnamefont {A.}~\bibnamefont
  {Donarini}}, \bibinfo {author} {\bibfnamefont {M.}~\bibnamefont {Niklas}},
  \bibinfo {author} {\bibfnamefont {M.}~\bibnamefont {Schafberger}}, \bibinfo
  {author} {\bibfnamefont {N.}~\bibnamefont {Paradiso}}, \bibinfo {author}
  {\bibfnamefont {C.}~\bibnamefont {Strunk}}, \ and\ \bibinfo {author}
  {\bibfnamefont {M.}~\bibnamefont {Grifoni}},\ }\href {\doibase
  10.1038/s41467-018-08112-x} {\bibfield  {journal} {\bibinfo  {journal}
  {Nature Communications}\ }\textbf {\bibinfo {volume} {10}},\ \bibinfo {pages}
  {381} (\bibinfo {year} {2019})}\BibitemShut {NoStop}%
\bibitem [{\citenamefont {Hills}\ \emph {et~al.}(2019)\citenamefont {Hills},
  \citenamefont {Lau}, \citenamefont {Wright}, \citenamefont {Fuller},
  \citenamefont {Bishop}, \citenamefont {Srimani}, \citenamefont {Kanhaiya},
  \citenamefont {Ho}, \citenamefont {Amer}, \citenamefont {Stein},
  \citenamefont {Murphy}, \citenamefont {Arvind}, \citenamefont
  {Chandrakasan},\ and\ \citenamefont {Shulaker}}]{Hills2019}%
  \BibitemOpen
  \bibfield  {author} {\bibinfo {author} {\bibfnamefont {G.}~\bibnamefont
  {Hills}}, \bibinfo {author} {\bibfnamefont {C.}~\bibnamefont {Lau}}, \bibinfo
  {author} {\bibfnamefont {A.}~\bibnamefont {Wright}}, \bibinfo {author}
  {\bibfnamefont {S.}~\bibnamefont {Fuller}}, \bibinfo {author} {\bibfnamefont
  {M.~D.}\ \bibnamefont {Bishop}}, \bibinfo {author} {\bibfnamefont
  {T.}~\bibnamefont {Srimani}}, \bibinfo {author} {\bibfnamefont
  {P.}~\bibnamefont {Kanhaiya}}, \bibinfo {author} {\bibfnamefont
  {R.}~\bibnamefont {Ho}}, \bibinfo {author} {\bibfnamefont {A.}~\bibnamefont
  {Amer}}, \bibinfo {author} {\bibfnamefont {Y.}~\bibnamefont {Stein}},
  \bibinfo {author} {\bibfnamefont {D.}~\bibnamefont {Murphy}}, \bibinfo
  {author} {\bibnamefont {Arvind}}, \bibinfo {author} {\bibfnamefont
  {A.}~\bibnamefont {Chandrakasan}}, \ and\ \bibinfo {author} {\bibfnamefont
  {M.~M.}\ \bibnamefont {Shulaker}},\ }\href {\doibase
  10.1038/s41586-019-1493-8} {\bibfield  {journal} {\bibinfo  {journal}
  {Nature}\ }\textbf {\bibinfo {volume} {572}},\ \bibinfo {pages} {595}
  (\bibinfo {year} {2019})}\BibitemShut {NoStop}%
\bibitem [{\citenamefont {Khivrich}\ \emph {et~al.}(2019)\citenamefont
  {Khivrich}, \citenamefont {Clerk},\ and\ \citenamefont
  {Ilani}}]{Khivrich2019}%
  \BibitemOpen
  \bibfield  {author} {\bibinfo {author} {\bibfnamefont {I.}~\bibnamefont
  {Khivrich}}, \bibinfo {author} {\bibfnamefont {A.~A.}\ \bibnamefont {Clerk}},
  \ and\ \bibinfo {author} {\bibfnamefont {S.}~\bibnamefont {Ilani}},\ }\href
  {\doibase 10.1038/s41565-018-0341-6} {\bibfield  {journal} {\bibinfo
  {journal} {Nature Nanotechnology}\ }\textbf {\bibinfo {volume} {14}},\
  \bibinfo {pages} {161} (\bibinfo {year} {2019})}\BibitemShut {NoStop}%
\bibitem [{\citenamefont {Shapir}\ \emph {et~al.}(2019)\citenamefont {Shapir},
  \citenamefont {Hamo}, \citenamefont {Pecker}, \citenamefont {Moca},
  \citenamefont {Legeza}, \citenamefont {Zarand},\ and\ \citenamefont
  {Ilani}}]{Shapir.2019}%
  \BibitemOpen
  \bibfield  {author} {\bibinfo {author} {\bibfnamefont {I.}~\bibnamefont
  {Shapir}}, \bibinfo {author} {\bibfnamefont {A.}~\bibnamefont {Hamo}},
  \bibinfo {author} {\bibfnamefont {S.}~\bibnamefont {Pecker}}, \bibinfo
  {author} {\bibfnamefont {C.~P.}\ \bibnamefont {Moca}}, \bibinfo {author}
  {\bibfnamefont {{\"O}.}~\bibnamefont {Legeza}}, \bibinfo {author}
  {\bibfnamefont {G.}~\bibnamefont {Zarand}}, \ and\ \bibinfo {author}
  {\bibfnamefont {S.}~\bibnamefont {Ilani}},\ }\href {\doibase
  10.1126/science.aat0905} {\bibfield  {journal} {\bibinfo  {journal}
  {Science}\ }\textbf {\bibinfo {volume} {364}},\ \bibinfo {pages} {870}
  (\bibinfo {year} {2019})}\BibitemShut {NoStop}%
\bibitem [{\citenamefont {Marga\ifmmode~\acute{n}\else \'{n}\fi{}ska}\ \emph
  {et~al.}(2019)\citenamefont {Marga\ifmmode~\acute{n}\else \'{n}\fi{}ska},
  \citenamefont {Schmid}, \citenamefont {Dirnaichner}, \citenamefont {Stiller},
  \citenamefont {Strunk}, \citenamefont {Grifoni},\ and\ \citenamefont
  {H\"uttel}}]{Grifoni.2019}%
  \BibitemOpen
  \bibfield  {author} {\bibinfo {author} {\bibfnamefont {M.}~\bibnamefont
  {Marga\ifmmode~\acute{n}\else \'{n}\fi{}ska}}, \bibinfo {author}
  {\bibfnamefont {D.~R.}\ \bibnamefont {Schmid}}, \bibinfo {author}
  {\bibfnamefont {A.}~\bibnamefont {Dirnaichner}}, \bibinfo {author}
  {\bibfnamefont {P.~L.}\ \bibnamefont {Stiller}}, \bibinfo {author}
  {\bibfnamefont {C.}~\bibnamefont {Strunk}}, \bibinfo {author} {\bibfnamefont
  {M.}~\bibnamefont {Grifoni}}, \ and\ \bibinfo {author} {\bibfnamefont
  {A.~K.}\ \bibnamefont {H\"uttel}},\ }\href {\doibase
  10.1103/PhysRevLett.122.086802} {\bibfield  {journal} {\bibinfo  {journal}
  {Phys. Rev. Lett.}\ }\textbf {\bibinfo {volume} {122}},\ \bibinfo {pages}
  {086802} (\bibinfo {year} {2019})}\BibitemShut {NoStop}%
\bibitem [{\citenamefont {Graf}\ \emph {et~al.}(2017)\citenamefont {Graf},
  \citenamefont {Held}, \citenamefont {Zakharko}, \citenamefont {Tropf},
  \citenamefont {Gather},\ and\ \citenamefont {Zaumseil}}]{Graf2017}%
  \BibitemOpen
  \bibfield  {author} {\bibinfo {author} {\bibfnamefont {A.}~\bibnamefont
  {Graf}}, \bibinfo {author} {\bibfnamefont {M.}~\bibnamefont {Held}}, \bibinfo
  {author} {\bibfnamefont {Y.}~\bibnamefont {Zakharko}}, \bibinfo {author}
  {\bibfnamefont {L.}~\bibnamefont {Tropf}}, \bibinfo {author} {\bibfnamefont
  {M.~C.}\ \bibnamefont {Gather}}, \ and\ \bibinfo {author} {\bibfnamefont
  {J.}~\bibnamefont {Zaumseil}},\ }\href {\doibase 10.1038/nmat4940} {\bibfield
   {journal} {\bibinfo  {journal} {Nature Materials}\ }\textbf {\bibinfo
  {volume} {16}},\ \bibinfo {pages} {911} (\bibinfo {year} {2017})}\BibitemShut
  {NoStop}%
\bibitem [{\citenamefont {Hata}\ \emph {et~al.}(2018)\citenamefont {Hata},
  \citenamefont {Delagrange}, \citenamefont {Arakawa}, \citenamefont {Lee},
  \citenamefont {Deblock}, \citenamefont {Bouchiat}, \citenamefont
  {Kobayashi},\ and\ \citenamefont {Ferrier}}]{Hata.2018}%
  \BibitemOpen
  \bibfield  {author} {\bibinfo {author} {\bibfnamefont {T.}~\bibnamefont
  {Hata}}, \bibinfo {author} {\bibfnamefont {R.}~\bibnamefont {Delagrange}},
  \bibinfo {author} {\bibfnamefont {T.}~\bibnamefont {Arakawa}}, \bibinfo
  {author} {\bibfnamefont {S.}~\bibnamefont {Lee}}, \bibinfo {author}
  {\bibfnamefont {R.}~\bibnamefont {Deblock}}, \bibinfo {author} {\bibfnamefont
  {H.}~\bibnamefont {Bouchiat}}, \bibinfo {author} {\bibfnamefont
  {K.}~\bibnamefont {Kobayashi}}, \ and\ \bibinfo {author} {\bibfnamefont
  {M.}~\bibnamefont {Ferrier}},\ }\href {\doibase
  10.1103/PhysRevLett.121.247703} {\bibfield  {journal} {\bibinfo  {journal}
  {Phys. Rev. Lett.}\ }\textbf {\bibinfo {volume} {121}},\ \bibinfo {pages}
  {247703} (\bibinfo {year} {2018})}\BibitemShut {NoStop}%
\bibitem [{\citenamefont {Efroni}\ \emph {et~al.}(2017)\citenamefont {Efroni},
  \citenamefont {Ilani},\ and\ \citenamefont {Berg}}]{Ilani}%
  \BibitemOpen
  \bibfield  {author} {\bibinfo {author} {\bibfnamefont {Y.}~\bibnamefont
  {Efroni}}, \bibinfo {author} {\bibfnamefont {S.}~\bibnamefont {Ilani}}, \
  and\ \bibinfo {author} {\bibfnamefont {E.}~\bibnamefont {Berg}},\ }\href
  {\doibase 10.1103/PhysRevLett.119.147704} {\bibfield  {journal} {\bibinfo
  {journal} {Phys. Rev. Lett.}\ }\textbf {\bibinfo {volume} {119}},\ \bibinfo
  {pages} {147704} (\bibinfo {year} {2017})}\BibitemShut {NoStop}%
\bibitem [{\citenamefont {Izumida}\ \emph {et~al.}(2016)\citenamefont
  {Izumida}, \citenamefont {Okuyama}, \citenamefont {Yamakage},\ and\
  \citenamefont {Saito}}]{Wataru.2016}%
  \BibitemOpen
  \bibfield  {author} {\bibinfo {author} {\bibfnamefont {W.}~\bibnamefont
  {Izumida}}, \bibinfo {author} {\bibfnamefont {R.}~\bibnamefont {Okuyama}},
  \bibinfo {author} {\bibfnamefont {A.}~\bibnamefont {Yamakage}}, \ and\
  \bibinfo {author} {\bibfnamefont {R.}~\bibnamefont {Saito}},\ }\href
  {\doibase 10.1103/PhysRevB.93.195442} {\bibfield  {journal} {\bibinfo
  {journal} {Phys. Rev. B}\ }\textbf {\bibinfo {volume} {93}},\ \bibinfo
  {pages} {195442} (\bibinfo {year} {2016})}\BibitemShut {NoStop}%
\bibitem [{\citenamefont {Okuyama}\ \emph {et~al.}(2019)\citenamefont
  {Okuyama}, \citenamefont {Izumida},\ and\ \citenamefont {Eto}}]{Wataru.2018}%
  \BibitemOpen
  \bibfield  {author} {\bibinfo {author} {\bibfnamefont {R.}~\bibnamefont
  {Okuyama}}, \bibinfo {author} {\bibfnamefont {W.}~\bibnamefont {Izumida}}, \
  and\ \bibinfo {author} {\bibfnamefont {M.}~\bibnamefont {Eto}},\ }\href
  {\doibase 10.1103/PhysRevB.99.115409} {\bibfield  {journal} {\bibinfo
  {journal} {Phys. Rev. B}\ }\textbf {\bibinfo {volume} {99}},\ \bibinfo
  {pages} {115409} (\bibinfo {year} {2019})}\BibitemShut {NoStop}%
\bibitem [{\citenamefont {Higuchi}\ \emph {et~al.}(2004)\citenamefont
  {Higuchi}, \citenamefont {Kusakabe}, \citenamefont {Suzuki}, \citenamefont
  {Tsuneyuki}, \citenamefont {Yamauchi}, \citenamefont {Akagi},\ and\
  \citenamefont {Yoshimoto}}]{Higuchi_2004}%
  \BibitemOpen
  \bibfield  {author} {\bibinfo {author} {\bibfnamefont {Y.}~\bibnamefont
  {Higuchi}}, \bibinfo {author} {\bibfnamefont {K.}~\bibnamefont {Kusakabe}},
  \bibinfo {author} {\bibfnamefont {N.}~\bibnamefont {Suzuki}}, \bibinfo
  {author} {\bibfnamefont {S.}~\bibnamefont {Tsuneyuki}}, \bibinfo {author}
  {\bibfnamefont {J.}~\bibnamefont {Yamauchi}}, \bibinfo {author}
  {\bibfnamefont {K.}~\bibnamefont {Akagi}}, \ and\ \bibinfo {author}
  {\bibfnamefont {Y.}~\bibnamefont {Yoshimoto}},\ }\href {\doibase
  10.1088/0953-8984/16/48/028} {\bibfield  {journal} {\bibinfo  {journal}
  {Journal of Physics: Condensed Matter}\ }\textbf {\bibinfo {volume} {16}},\
  \bibinfo {pages} {S5689} (\bibinfo {year} {2004})}\BibitemShut {NoStop}%
\bibitem [{\citenamefont {Hod}\ and\ \citenamefont {Scuseria}(2008)}]{Hod2008}%
  \BibitemOpen
  \bibfield  {author} {\bibinfo {author} {\bibfnamefont {O.}~\bibnamefont
  {Hod}}\ and\ \bibinfo {author} {\bibfnamefont {G.~E.}\ \bibnamefont
  {Scuseria}},\ }\href {\doibase 10.1021/nn8004069} {\bibfield  {journal}
  {\bibinfo  {journal} {ACS Nano}\ }\textbf {\bibinfo {volume} {2}},\ \bibinfo
  {pages} {2243} (\bibinfo {year} {2008})}\BibitemShut {NoStop}%
\bibitem [{\citenamefont {Lin}(1998)}]{Lin}%
  \BibitemOpen
  \bibfield  {author} {\bibinfo {author} {\bibfnamefont {H.-H.}\ \bibnamefont
  {Lin}},\ }\href {\doibase 10.1103/PhysRevB.58.4963} {\bibfield  {journal}
  {\bibinfo  {journal} {Phys. Rev. B}\ }\textbf {\bibinfo {volume} {58}},\
  \bibinfo {pages} {4963} (\bibinfo {year} {1998})}\BibitemShut {NoStop}%
\bibitem [{\citenamefont {Ma\~nanes}\ \emph {et~al.}(2008)\citenamefont
  {Ma\~nanes}, \citenamefont {Duque}, \citenamefont {Ayuela}, \citenamefont
  {L\'opez},\ and\ \citenamefont {Alonso}}]{Mananes}%
  \BibitemOpen
  \bibfield  {author} {\bibinfo {author} {\bibfnamefont {A.}~\bibnamefont
  {Ma\~nanes}}, \bibinfo {author} {\bibfnamefont {F.}~\bibnamefont {Duque}},
  \bibinfo {author} {\bibfnamefont {A.}~\bibnamefont {Ayuela}}, \bibinfo
  {author} {\bibfnamefont {M.~J.}\ \bibnamefont {L\'opez}}, \ and\ \bibinfo
  {author} {\bibfnamefont {J.~A.}\ \bibnamefont {Alonso}},\ }\href {\doibase
  10.1103/PhysRevB.78.035432} {\bibfield  {journal} {\bibinfo  {journal} {Phys.
  Rev. B}\ }\textbf {\bibinfo {volume} {78}},\ \bibinfo {pages} {035432}
  (\bibinfo {year} {2008})}\BibitemShut {NoStop}%
\bibitem [{\citenamefont {Kunstmann}\ \emph {et~al.}(2011)\citenamefont
  {Kunstmann}, \citenamefont {\"Ozdo\ifmmode~\breve{g}\else \u{g}\fi{}an},
  \citenamefont {Quandt},\ and\ \citenamefont {Fehske}}]{Kunstmann}%
  \BibitemOpen
  \bibfield  {author} {\bibinfo {author} {\bibfnamefont {J.}~\bibnamefont
  {Kunstmann}}, \bibinfo {author} {\bibfnamefont {C.}~\bibnamefont
  {\"Ozdo\ifmmode~\breve{g}\else \u{g}\fi{}an}}, \bibinfo {author}
  {\bibfnamefont {A.}~\bibnamefont {Quandt}}, \ and\ \bibinfo {author}
  {\bibfnamefont {H.}~\bibnamefont {Fehske}},\ }\href {\doibase
  10.1103/PhysRevB.83.045414} {\bibfield  {journal} {\bibinfo  {journal} {Phys.
  Rev. B}\ }\textbf {\bibinfo {volume} {83}},\ \bibinfo {pages} {045414}
  (\bibinfo {year} {2011})}\BibitemShut {NoStop}%
\bibitem [{\citenamefont {Wakabayashi}\ \emph {et~al.}(2010)\citenamefont
  {Wakabayashi}, \citenamefont {Sasaki}, \citenamefont {Nakanishi},\ and\
  \citenamefont {Enoki}}]{Wakabayashi_2010}%
  \BibitemOpen
  \bibfield  {author} {\bibinfo {author} {\bibfnamefont {K.}~\bibnamefont
  {Wakabayashi}}, \bibinfo {author} {\bibfnamefont {K.}~\bibnamefont {Sasaki}},
  \bibinfo {author} {\bibfnamefont {T.}~\bibnamefont {Nakanishi}}, \ and\
  \bibinfo {author} {\bibfnamefont {T.}~\bibnamefont {Enoki}},\ }\href
  {\doibase 10.1088/1468-6996/11/5/054504} {\bibfield  {journal} {\bibinfo
  {journal} {Science and Technology of Advanced Materials}\ }\textbf {\bibinfo
  {volume} {11}},\ \bibinfo {pages} {054504} (\bibinfo {year}
  {2010})}\BibitemShut {NoStop}%
\bibitem [{\citenamefont {Dutta}\ and\ \citenamefont
  {Wakabayashi}(2012)}]{Dutta2012}%
  \BibitemOpen
  \bibfield  {author} {\bibinfo {author} {\bibfnamefont {S.}~\bibnamefont
  {Dutta}}\ and\ \bibinfo {author} {\bibfnamefont {K.}~\bibnamefont
  {Wakabayashi}},\ }\href {https://doi.org/10.1038/srep00519} {\bibfield
  {journal} {\bibinfo  {journal} {Scientific Reports}\ }\textbf {\bibinfo
  {volume} {2}},\ \bibinfo {pages} {519} (\bibinfo {year} {2012})}\BibitemShut
  {NoStop}%
\bibitem [{\citenamefont {Yazyev}\ \emph {et~al.}(2011)\citenamefont {Yazyev},
  \citenamefont {Capaz},\ and\ \citenamefont {Louie}}]{Yazyev}%
  \BibitemOpen
  \bibfield  {author} {\bibinfo {author} {\bibfnamefont {O.~V.}\ \bibnamefont
  {Yazyev}}, \bibinfo {author} {\bibfnamefont {R.~B.}\ \bibnamefont {Capaz}}, \
  and\ \bibinfo {author} {\bibfnamefont {S.~G.}\ \bibnamefont {Louie}},\ }\href
  {\doibase 10.1103/PhysRevB.84.115406} {\bibfield  {journal} {\bibinfo
  {journal} {Phys. Rev. B}\ }\textbf {\bibinfo {volume} {84}},\ \bibinfo
  {pages} {115406} (\bibinfo {year} {2011})}\BibitemShut {NoStop}%
\bibitem [{\citenamefont {Magda}\ \emph {et~al.}(2014)\citenamefont {Magda},
  \citenamefont {Jin}, \citenamefont {Hagym{\'a}si}, \citenamefont
  {Vancs{\'o}}, \citenamefont {Osv{\'a}th}, \citenamefont {Nemes-Incze},
  \citenamefont {Hwang}, \citenamefont {Bir{\'o}},\ and\ \citenamefont
  {Tapaszt{\'o}}}]{Magda2014}%
  \BibitemOpen
  \bibfield  {author} {\bibinfo {author} {\bibfnamefont {G.~Z.}\ \bibnamefont
  {Magda}}, \bibinfo {author} {\bibfnamefont {X.}~\bibnamefont {Jin}}, \bibinfo
  {author} {\bibfnamefont {I.}~\bibnamefont {Hagym{\'a}si}}, \bibinfo {author}
  {\bibfnamefont {P.}~\bibnamefont {Vancs{\'o}}}, \bibinfo {author}
  {\bibfnamefont {Z.}~\bibnamefont {Osv{\'a}th}}, \bibinfo {author}
  {\bibfnamefont {P.}~\bibnamefont {Nemes-Incze}}, \bibinfo {author}
  {\bibfnamefont {C.}~\bibnamefont {Hwang}}, \bibinfo {author} {\bibfnamefont
  {L.~P.}\ \bibnamefont {Bir{\'o}}}, \ and\ \bibinfo {author} {\bibfnamefont
  {L.}~\bibnamefont {Tapaszt{\'o}}},\ }\href
  {https://doi.org/10.1038/nature13831} {\bibfield  {journal} {\bibinfo
  {journal} {Nature}\ }\textbf {\bibinfo {volume} {514}},\ \bibinfo {pages}
  {608} (\bibinfo {year} {2014})}\BibitemShut {NoStop}%
\bibitem [{\citenamefont {Tao}\ \emph {et~al.}(2011)\citenamefont {Tao},
  \citenamefont {Jiao}, \citenamefont {Yazyev}, \citenamefont {Chen},
  \citenamefont {Feng}, \citenamefont {Zhang}, \citenamefont {Capaz},
  \citenamefont {Tour}, \citenamefont {Zettl}, \citenamefont {Louie},
  \citenamefont {Dai},\ and\ \citenamefont {Crommie}}]{Tao2011}%
  \BibitemOpen
  \bibfield  {author} {\bibinfo {author} {\bibfnamefont {C.}~\bibnamefont
  {Tao}}, \bibinfo {author} {\bibfnamefont {L.}~\bibnamefont {Jiao}}, \bibinfo
  {author} {\bibfnamefont {O.~V.}\ \bibnamefont {Yazyev}}, \bibinfo {author}
  {\bibfnamefont {Y.-C.}\ \bibnamefont {Chen}}, \bibinfo {author}
  {\bibfnamefont {J.}~\bibnamefont {Feng}}, \bibinfo {author} {\bibfnamefont
  {X.}~\bibnamefont {Zhang}}, \bibinfo {author} {\bibfnamefont {R.~B.}\
  \bibnamefont {Capaz}}, \bibinfo {author} {\bibfnamefont {J.~M.}\ \bibnamefont
  {Tour}}, \bibinfo {author} {\bibfnamefont {A.}~\bibnamefont {Zettl}},
  \bibinfo {author} {\bibfnamefont {S.~G.}\ \bibnamefont {Louie}}, \bibinfo
  {author} {\bibfnamefont {H.}~\bibnamefont {Dai}}, \ and\ \bibinfo {author}
  {\bibfnamefont {M.~F.}\ \bibnamefont {Crommie}},\ }\href
  {https://doi.org/10.1038/nphys1991} {\bibfield  {journal} {\bibinfo
  {journal} {Nature Physics}\ }\textbf {\bibinfo {volume} {7}},\ \bibinfo
  {pages} {616} (\bibinfo {year} {2011})}\BibitemShut {NoStop}%
\bibitem [{\citenamefont {Reich}\ \emph {et~al.}(2002)\citenamefont {Reich},
  \citenamefont {Maultzsch}, \citenamefont {Thomsen},\ and\ \citenamefont
  {Ordej\'on}}]{Reich}%
  \BibitemOpen
  \bibfield  {author} {\bibinfo {author} {\bibfnamefont {S.}~\bibnamefont
  {Reich}}, \bibinfo {author} {\bibfnamefont {J.}~\bibnamefont {Maultzsch}},
  \bibinfo {author} {\bibfnamefont {C.}~\bibnamefont {Thomsen}}, \ and\
  \bibinfo {author} {\bibfnamefont {P.}~\bibnamefont {Ordej\'on}},\ }\href
  {\doibase 10.1103/PhysRevB.66.035412} {\bibfield  {journal} {\bibinfo
  {journal} {Phys. Rev. B}\ }\textbf {\bibinfo {volume} {66}},\ \bibinfo
  {pages} {035412} (\bibinfo {year} {2002})}\BibitemShut {NoStop}%
\bibitem [{\citenamefont {Laird}\ \emph
  {et~al.}(2015{\natexlab{b}})\citenamefont {Laird}, \citenamefont {Kuemmeth},
  \citenamefont {Steele}, \citenamefont {Grove-Rasmussen}, \citenamefont
  {Nyg\aa{}rd}, \citenamefont {Flensberg},\ and\ \citenamefont
  {Kouwenhoven}}]{KouwenhovenRMP}%
  \BibitemOpen
  \bibfield  {author} {\bibinfo {author} {\bibfnamefont {E.~A.}\ \bibnamefont
  {Laird}}, \bibinfo {author} {\bibfnamefont {F.}~\bibnamefont {Kuemmeth}},
  \bibinfo {author} {\bibfnamefont {G.~A.}\ \bibnamefont {Steele}}, \bibinfo
  {author} {\bibfnamefont {K.}~\bibnamefont {Grove-Rasmussen}}, \bibinfo
  {author} {\bibfnamefont {J.}~\bibnamefont {Nyg\aa{}rd}}, \bibinfo {author}
  {\bibfnamefont {K.}~\bibnamefont {Flensberg}}, \ and\ \bibinfo {author}
  {\bibfnamefont {L.~P.}\ \bibnamefont {Kouwenhoven}},\ }\href {\doibase
  10.1103/RevModPhys.87.703} {\bibfield  {journal} {\bibinfo  {journal} {Rev.
  Mod. Phys.}\ }\textbf {\bibinfo {volume} {87}},\ \bibinfo {pages} {703}
  (\bibinfo {year} {2015}{\natexlab{b}})}\BibitemShut {NoStop}%
\bibitem [{\citenamefont {Kuemmeth}\ \emph {et~al.}(2008)\citenamefont
  {Kuemmeth}, \citenamefont {Ilani}, \citenamefont {Ralph},\ and\ \citenamefont
  {McEuen}}]{Kuemmeth2008}%
  \BibitemOpen
  \bibfield  {author} {\bibinfo {author} {\bibfnamefont {F.}~\bibnamefont
  {Kuemmeth}}, \bibinfo {author} {\bibfnamefont {S.}~\bibnamefont {Ilani}},
  \bibinfo {author} {\bibfnamefont {D.~C.}\ \bibnamefont {Ralph}}, \ and\
  \bibinfo {author} {\bibfnamefont {P.~L.}\ \bibnamefont {McEuen}},\ }\href
  {\doibase 10.1038/nature06822} {\bibfield  {journal} {\bibinfo  {journal}
  {Nature}\ }\textbf {\bibinfo {volume} {452}},\ \bibinfo {pages} {448}
  (\bibinfo {year} {2008})}\BibitemShut {NoStop}%
\bibitem [{\citenamefont {Perebeinos}\ \emph {et~al.}(2004)\citenamefont
  {Perebeinos}, \citenamefont {Tersoff},\ and\ \citenamefont
  {Avouris}}]{Perebeinos-PRL-2004}%
  \BibitemOpen
  \bibfield  {author} {\bibinfo {author} {\bibfnamefont {V.}~\bibnamefont
  {Perebeinos}}, \bibinfo {author} {\bibfnamefont {J.}~\bibnamefont {Tersoff}},
  \ and\ \bibinfo {author} {\bibfnamefont {P.}~\bibnamefont {Avouris}},\ }\href
  {\doibase 10.1103/PhysRevLett.92.257402} {\bibfield  {journal} {\bibinfo
  {journal} {Phys. Rev. Lett.}\ }\textbf {\bibinfo {volume} {92}},\ \bibinfo
  {pages} {257402} (\bibinfo {year} {2004})}\BibitemShut {NoStop}%
\bibitem [{\citenamefont {White}\ \emph {et~al.}(1993)\citenamefont {White},
  \citenamefont {Robertson},\ and\ \citenamefont {Mintmire}}]{White}%
  \BibitemOpen
  \bibfield  {author} {\bibinfo {author} {\bibfnamefont {C.~T.}\ \bibnamefont
  {White}}, \bibinfo {author} {\bibfnamefont {D.~H.}\ \bibnamefont
  {Robertson}}, \ and\ \bibinfo {author} {\bibfnamefont {J.~W.}\ \bibnamefont
  {Mintmire}},\ }\href {\doibase 10.1103/PhysRevB.47.5485} {\bibfield
  {journal} {\bibinfo  {journal} {Phys. Rev. B}\ }\textbf {\bibinfo {volume}
  {47}},\ \bibinfo {pages} {5485} (\bibinfo {year} {1993})}\BibitemShut
  {NoStop}%
\bibitem [{\citenamefont {Jishi}\ \emph {et~al.}(1993)\citenamefont {Jishi},
  \citenamefont {Dresselhaus},\ and\ \citenamefont {Dresselhaus}}]{Jishi}%
  \BibitemOpen
  \bibfield  {author} {\bibinfo {author} {\bibfnamefont {R.~A.}\ \bibnamefont
  {Jishi}}, \bibinfo {author} {\bibfnamefont {M.~S.}\ \bibnamefont
  {Dresselhaus}}, \ and\ \bibinfo {author} {\bibfnamefont {G.}~\bibnamefont
  {Dresselhaus}},\ }\href {\doibase 10.1103/PhysRevB.47.16671} {\bibfield
  {journal} {\bibinfo  {journal} {Phys. Rev. B}\ }\textbf {\bibinfo {volume}
  {47}},\ \bibinfo {pages} {16671} (\bibinfo {year} {1993})}\BibitemShut
  {NoStop}%
\bibitem [{Note1()}]{Note1}%
  \BibitemOpen
  \bibinfo {note} {See Supplementary material for details on the effective 1D
  lattice model and DMRG calculations which includes Refs.~\cite {Reich, White,
  Jishi,KouwenhovenRMP,Perebeinos-PRL-2004}}\BibitemShut {NoStop}%
\bibitem [{\citenamefont {Hatsugai}(1993)}]{Hatsugai}%
  \BibitemOpen
  \bibfield  {author} {\bibinfo {author} {\bibfnamefont {Y.}~\bibnamefont
  {Hatsugai}},\ }\href {\doibase 10.1103/PhysRevLett.71.3697} {\bibfield
  {journal} {\bibinfo  {journal} {Phys. Rev. Lett.}\ }\textbf {\bibinfo
  {volume} {71}},\ \bibinfo {pages} {3697} (\bibinfo {year}
  {1993})}\BibitemShut {NoStop}%
\bibitem [{\citenamefont {Teo}\ and\ \citenamefont
  {Kane}(2010)}]{teo2010topological}%
  \BibitemOpen
  \bibfield  {author} {\bibinfo {author} {\bibfnamefont {J.~C.}\ \bibnamefont
  {Teo}}\ and\ \bibinfo {author} {\bibfnamefont {C.~L.}\ \bibnamefont {Kane}},\
  }\href@noop {} {\bibfield  {journal} {\bibinfo  {journal} {Physical Review
  B}\ }\textbf {\bibinfo {volume} {82}},\ \bibinfo {pages} {115120} (\bibinfo
  {year} {2010})}\BibitemShut {NoStop}%
\bibitem [{\citenamefont {Akhmerov}\ and\ \citenamefont
  {Beenakker}(2008)}]{akhmerov2008boundary}%
  \BibitemOpen
  \bibfield  {author} {\bibinfo {author} {\bibfnamefont {A.}~\bibnamefont
  {Akhmerov}}\ and\ \bibinfo {author} {\bibfnamefont {C.}~\bibnamefont
  {Beenakker}},\ }\href@noop {} {\bibfield  {journal} {\bibinfo  {journal}
  {Physical Review B}\ }\textbf {\bibinfo {volume} {77}},\ \bibinfo {pages}
  {085423} (\bibinfo {year} {2008})}\BibitemShut {NoStop}%
\bibitem [{\citenamefont {White}(1996)}]{WhiteDMRG}%
  \BibitemOpen
  \bibfield  {author} {\bibinfo {author} {\bibfnamefont {S.~R.}\ \bibnamefont
  {White}},\ }\href {\doibase 10.1103/PhysRevLett.77.3633} {\bibfield
  {journal} {\bibinfo  {journal} {Phys. Rev. Lett.}\ }\textbf {\bibinfo
  {volume} {77}},\ \bibinfo {pages} {3633} (\bibinfo {year}
  {1996})}\BibitemShut {NoStop}%
\bibitem [{\citenamefont {Schollw\"ock}(2005)}]{Schollwock}%
  \BibitemOpen
  \bibfield  {author} {\bibinfo {author} {\bibfnamefont {U.}~\bibnamefont
  {Schollw\"ock}},\ }\href {\doibase 10.1103/RevModPhys.77.259} {\bibfield
  {journal} {\bibinfo  {journal} {Rev. Mod. Phys.}\ }\textbf {\bibinfo {volume}
  {77}},\ \bibinfo {pages} {259} (\bibinfo {year} {2005})}\BibitemShut
  {NoStop}%
\bibitem [{\citenamefont {Legeza}\ and\ \citenamefont {S\'olyom}(2003)}]{Ors}%
  \BibitemOpen
  \bibfield  {author} {\bibinfo {author} {\bibfnamefont {{\"O}.}~\bibnamefont
  {Legeza}}\ and\ \bibinfo {author} {\bibfnamefont {J.}~\bibnamefont
  {S\'olyom}},\ }\href {\doibase 10.1103/PhysRevB.68.195116} {\bibfield
  {journal} {\bibinfo  {journal} {Phys. Rev. B}\ }\textbf {\bibinfo {volume}
  {68}},\ \bibinfo {pages} {195116} (\bibinfo {year} {2003})}\BibitemShut
  {NoStop}%
\bibitem [{\citenamefont {Krumnow}\ \emph {et~al.}(2016)\citenamefont
  {Krumnow}, \citenamefont {Veis}, \citenamefont {Legeza},\ and\ \citenamefont
  {Eisert}}]{Krumnow}%
  \BibitemOpen
  \bibfield  {author} {\bibinfo {author} {\bibfnamefont {C.}~\bibnamefont
  {Krumnow}}, \bibinfo {author} {\bibfnamefont {L.}~\bibnamefont {Veis}},
  \bibinfo {author} {\bibfnamefont {{\"O}.}~\bibnamefont {Legeza}}, \ and\
  \bibinfo {author} {\bibfnamefont {J.}~\bibnamefont {Eisert}},\ }\href
  {\doibase 10.1103/PhysRevLett.117.210402} {\bibfield  {journal} {\bibinfo
  {journal} {Phys. Rev. Lett.}\ }\textbf {\bibinfo {volume} {117}},\ \bibinfo
  {pages} {210402} (\bibinfo {year} {2016})}\BibitemShut {NoStop}%
\bibitem [{\citenamefont {Simon}\ \emph {et~al.}(2006)\citenamefont {Simon},
  \citenamefont {Kuzmany}, \citenamefont {N\'afr\'adi}, \citenamefont
  {Feh\'er}, \citenamefont {Forr\'o}, \citenamefont {F\"ul\"op}, \citenamefont
  {J\'anossy}, \citenamefont {Korecz}, \citenamefont {Rockenbauer},
  \citenamefont {Hauke},\ and\ \citenamefont {Hirsch}}]{Ferenc}%
  \BibitemOpen
  \bibfield  {author} {\bibinfo {author} {\bibfnamefont {F.}~\bibnamefont
  {Simon}}, \bibinfo {author} {\bibfnamefont {H.}~\bibnamefont {Kuzmany}},
  \bibinfo {author} {\bibfnamefont {B.}~\bibnamefont {N\'afr\'adi}}, \bibinfo
  {author} {\bibfnamefont {T.}~\bibnamefont {Feh\'er}}, \bibinfo {author}
  {\bibfnamefont {L.}~\bibnamefont {Forr\'o}}, \bibinfo {author} {\bibfnamefont
  {F.}~\bibnamefont {F\"ul\"op}}, \bibinfo {author} {\bibfnamefont
  {A.}~\bibnamefont {J\'anossy}}, \bibinfo {author} {\bibfnamefont
  {L.}~\bibnamefont {Korecz}}, \bibinfo {author} {\bibfnamefont
  {A.}~\bibnamefont {Rockenbauer}}, \bibinfo {author} {\bibfnamefont
  {F.}~\bibnamefont {Hauke}}, \ and\ \bibinfo {author} {\bibfnamefont
  {A.}~\bibnamefont {Hirsch}},\ }\href {\doibase 10.1103/PhysRevLett.97.136801}
  {\bibfield  {journal} {\bibinfo  {journal} {Phys. Rev. Lett.}\ }\textbf
  {\bibinfo {volume} {97}},\ \bibinfo {pages} {136801} (\bibinfo {year}
  {2006})}\BibitemShut {NoStop}%
\bibitem [{\citenamefont {Náfrádi}\ \emph {et~al.}(2006)\citenamefont
  {Náfrádi}, \citenamefont {Nemes}, \citenamefont {Fehér}, \citenamefont
  {Forró}, \citenamefont {Kim}, \citenamefont {Fischer}, \citenamefont
  {Luzzi}, \citenamefont {Simon},\ and\ \citenamefont {Kuzmany}}]{nafradi}%
  \BibitemOpen
  \bibfield  {author} {\bibinfo {author} {\bibfnamefont {B.}~\bibnamefont
  {Náfrádi}}, \bibinfo {author} {\bibfnamefont {N.~M.}\ \bibnamefont
  {Nemes}}, \bibinfo {author} {\bibfnamefont {T.}~\bibnamefont {Fehér}},
  \bibinfo {author} {\bibfnamefont {L.}~\bibnamefont {Forró}}, \bibinfo
  {author} {\bibfnamefont {Y.}~\bibnamefont {Kim}}, \bibinfo {author}
  {\bibfnamefont {J.~E.}\ \bibnamefont {Fischer}}, \bibinfo {author}
  {\bibfnamefont {D.~E.}\ \bibnamefont {Luzzi}}, \bibinfo {author}
  {\bibfnamefont {F.}~\bibnamefont {Simon}}, \ and\ \bibinfo {author}
  {\bibfnamefont {H.}~\bibnamefont {Kuzmany}},\ }\href {\doibase
  10.1002/pssb.200669218} {\bibfield  {journal} {\bibinfo  {journal} {Phys.
  Stat. Sol. (b)}\ }\textbf {\bibinfo {volume} {243}},\ \bibinfo {pages} {3106}
  (\bibinfo {year} {2006})}\BibitemShut {NoStop}%
\bibitem [{\citenamefont {Hagym\'asi}\ and\ \citenamefont
  {Legeza}(2016)}]{Ors2016}%
  \BibitemOpen
  \bibfield  {author} {\bibinfo {author} {\bibfnamefont {I.}~\bibnamefont
  {Hagym\'asi}}\ and\ \bibinfo {author} {\bibfnamefont {O.}~\bibnamefont
  {Legeza}},\ }\href {\doibase 10.1103/PhysRevB.94.165147} {\bibfield
  {journal} {\bibinfo  {journal} {Phys. Rev. B}\ }\textbf {\bibinfo {volume}
  {94}},\ \bibinfo {pages} {165147} (\bibinfo {year} {2016})}\BibitemShut
  {NoStop}%
\bibitem [{\citenamefont {Wang}\ \emph {et~al.}(2005)\citenamefont {Wang},
  \citenamefont {Dukovic}, \citenamefont {Brus},\ and\ \citenamefont
  {Heinz}}]{Wang.2005}%
  \BibitemOpen
  \bibfield  {author} {\bibinfo {author} {\bibfnamefont {F.}~\bibnamefont
  {Wang}}, \bibinfo {author} {\bibfnamefont {G.}~\bibnamefont {Dukovic}},
  \bibinfo {author} {\bibfnamefont {L.~E.}\ \bibnamefont {Brus}}, \ and\
  \bibinfo {author} {\bibfnamefont {T.~F.}\ \bibnamefont {Heinz}},\ }\href
  {\doibase 10.1126/science.1110265} {\bibfield  {journal} {\bibinfo  {journal}
  {Science}\ }\textbf {\bibinfo {volume} {308}},\ \bibinfo {pages} {838}
  (\bibinfo {year} {2005})}\BibitemShut {NoStop}%
\bibitem [{\citenamefont {Maultzsch}\ \emph {et~al.}(2005)\citenamefont
  {Maultzsch}, \citenamefont {Pomraenke}, \citenamefont {Reich}, \citenamefont
  {Chang}, \citenamefont {Prezzi}, \citenamefont {Ruini}, \citenamefont
  {Molinari}, \citenamefont {Strano}, \citenamefont {Thomsen},\ and\
  \citenamefont {Lienau}}]{Maultzsch.2005}%
  \BibitemOpen
  \bibfield  {author} {\bibinfo {author} {\bibfnamefont {J.}~\bibnamefont
  {Maultzsch}}, \bibinfo {author} {\bibfnamefont {R.}~\bibnamefont
  {Pomraenke}}, \bibinfo {author} {\bibfnamefont {S.}~\bibnamefont {Reich}},
  \bibinfo {author} {\bibfnamefont {E.}~\bibnamefont {Chang}}, \bibinfo
  {author} {\bibfnamefont {D.}~\bibnamefont {Prezzi}}, \bibinfo {author}
  {\bibfnamefont {A.}~\bibnamefont {Ruini}}, \bibinfo {author} {\bibfnamefont
  {E.}~\bibnamefont {Molinari}}, \bibinfo {author} {\bibfnamefont {M.~S.}\
  \bibnamefont {Strano}}, \bibinfo {author} {\bibfnamefont {C.}~\bibnamefont
  {Thomsen}}, \ and\ \bibinfo {author} {\bibfnamefont {C.}~\bibnamefont
  {Lienau}},\ }\href {\doibase 10.1103/PhysRevB.72.241402} {\bibfield
  {journal} {\bibinfo  {journal} {Phys. Rev. B}\ }\textbf {\bibinfo {volume}
  {72}},\ \bibinfo {pages} {241402} (\bibinfo {year} {2005})}\BibitemShut
  {NoStop}%
\end{thebibliography}%
\clearpage

\section{Supplemental Information for "Topologically protected, correlated end spin formation in carbon nanotubes"}
        
\author{C\u at\u alin Pa\c scu Moca}
\affiliation{MTA-BME Quantum Dynamics and Correlations Research Group, 
Institute of Physics, Budapest University of Technology and Economics, 
Budafoki \'ut 8., H-1111 Budapest, Hungary}
\affiliation{Department of Physics, University of Oradea, 410087, Oradea, Romania}
\author{Wataru Izumida}
\affiliation{Department of Physics, Tohoku University, Sendai 980-8578, Japan}
\author{Bal\' azs D\' ora}
\affiliation{Department of Theoretical Physics and MTA-BME Lend\"ulet Topology and Correlation Research Group,
Budapest University of Technology and Economics, 1521 Budapest, Hungary}
\author{\" Ors Legeza}
\affiliation{Strongly Correlated Systems Lend\" ulet Research Group, Institute for Solid State Physics and Optics,
 MTA Wigner Research Centre for Physics, P.O. Box 49, H-1525 Budapest, Hungary}	
\author{Gergely Zar\'and}
\affiliation{MTA-BME Quantum Dynamics and Correlations Research Group, 
Institute of Physics, Budapest University of Technology and Economics, 
Budafoki \'ut 8., H-1111 Budapest, Hungary}
\affiliation{BME-MTA Exotic Quantum Phases 'Lend\"ulet' Research Group, Institute of Physics, Budapest University of Technology and Economics, 
Budafoki \'ut 8., H-1111 Budapest, Hungary}
\date{\today}
\maketitle
\subsection{ Effective 1D lattice model} 
\begin{figure}[b!]
 \includegraphics[width=0.9\columnwidth]{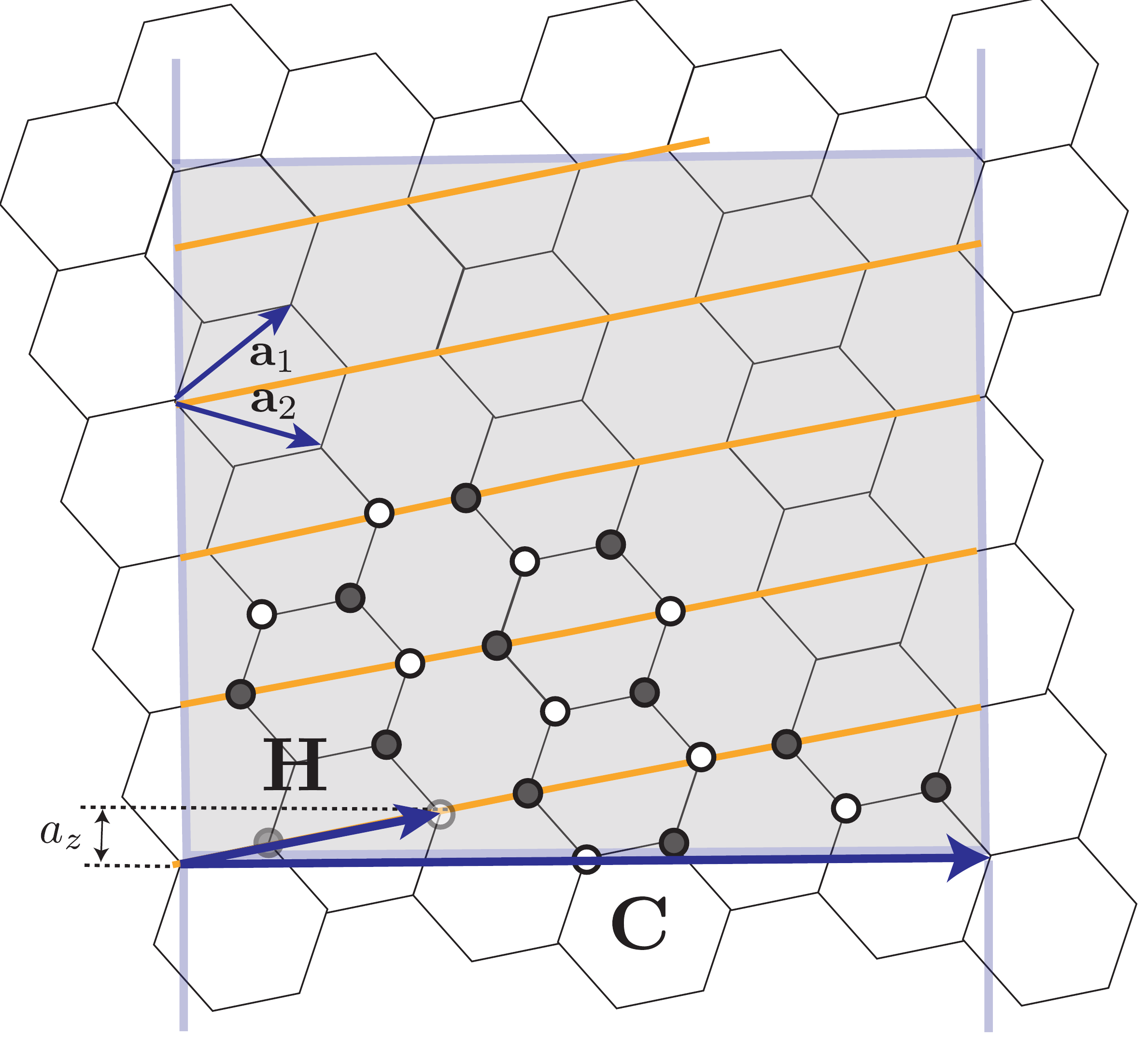}
  \caption{
    Hexagonal lattice structure and helical labeling of atoms for a $\chi=(4,2)$ nanotube.
    The chirality vector ${\bf C} = 4 {\bf a}_1 + 2 {\bf a}_2$  defines the direction, along which the 
    nanotube is rolled up. Atoms aligned along the  helicity vector,  ${\bf H} =  {\bf a}_1 +  {\bf a}_2$  
   form a helix, indicated by the orange line. The conventional unit cell is indicated as a gray rectangle, 
   its vertical edge is given by ${\bf T}= -4  {\bf a}_1 +  5 {\bf a}_2 $. 
   Since $d=\mathrm{gcd}(n,m)=2$ in this case, the vector $ {\bf C}/2$ brings the graphene sheet into 
   itself, i.e., the $(4,2)$ nanotube possesses a $C_2$ rotational symmetry. Correspondingly, one can 
   make two helixes of atoms $A$ and two of atoms $B$.}
  \label{fig:effective_model}
\end{figure}

In this section, we describe how to construct the effective Hamiltonian.  
Our starting point is the quadratic tight binding Hamiltonian describing the underlying graphene sheet
\begin{align}
  H_0
  = 
   -\sum_{ \bx, \bx', s} 
   t(\bx-\bx')
  c^\dagger _{s}(\bx)c_{s}(\bx')  \,,
   \label{eq:H_0}
\end{align}
where $c_s(\bx)$ denotes  the annihilation operator of an electron at site $\bx$ and  with spin
 $s = \{\uparrow, \downarrow\}$,  while  $t(\bx-\bx')$ represents the hopping integral between  
 lattice sites $\bx$ and $\bx'$. In our calculations, we also include second nearest neighbor hoppings to 
account for curvature effects~\cite{Reich} (see  Fig.~\ref{fig:effective_model}(b)).  

A single wall carbon nanotube  is  characterized by its chirality $\chi =(n,m)$, 
where $n$ and $m$ are integers. These  specify   the chirality vector, $\bC$, 
and the helical vector, $\bH$, as~\cite{White, Jishi}
\beq
\bC = n\,\ba_1+m\,\ba_2\, ,\hspace{1cm}
\bH = p\,\ba_1+q\,\ba_2\, .
\label{eq:vectors}
\eeq
Here $p$ and $q$ are two integers that satisfy the relation
$ m p - n q = d$, with $d=\mathrm{gcd}(n,m)$ the greatest common divisor of $n$ and $m$. 

The nanotube is obtained by rolling up a  graphene sheet along  $\bC$ (see Fig.~\ref{fig:effective_model}). As explained in the main text, the rolled up 
tube is invariant under $C_d$ rotations (corresponding to translations by a lattice vector $\bC/d$ of  the graphene sheet) 
as well as under  gliding rotations along the tube (generated by translations with $\bH$).
As illustrated in Fig.~\ref{fig:effective_model}, atoms 
forming the tube can correspondingly be organized into $d$ helixes of atoms $A$ and $d$ helixes of atoms $B$. 

Atoms on the nanotube are located at some physical positions $\br$. These positions 
can easily be expressed
 in terms of the atoms \emph{original} position  $\bx$ before the roll-up, expressed as
\be
  \bx = 
  \begin{cases}
   \frac{\nu}{d} \,\bC   + \ell\, \bH \, & \text{if $\bf$ in $A$},
\\
 \frac{\nu}{d}\, \bC   + \ell\, \bH + \frac{ \ba_1 + \ba_2} {3}
  \, &\text{if $\bf$ in $B$,}
  \end{cases}
  \label{eq:r_atomicPosition}
\ee
with the integer  $\nu =\{ 0, 1, \cdots, d-1\}$ specifying the helix, and $\ell$ the location along this helix. 
Notice that periodic boundary conditions are used within the graphene plane, i.e., 
graphene atoms with coordinates $\bx$ and $\bx+\bC$ are considered to be identical, and 
neighbors are identified accordingly. In our numerics, we use this helical construction, i.e., we specify atoms 
on the nanotube by the quantum numbers $\nu$, $\ell$, and the sublattice label $\tau  = A$ or $B$. 

The integer label $\ell$ can also be thought of as an indicator of  the lattice position along the nanotube
in units of
\begin{equation}
 a_z = \frac{T}{N/d}\, ,
\end{equation}
which is the shortest distance between two consecutive atoms along one helix, as projected to 
the axial direction (see Fig.~\ref{fig:effective_model}(b)).
Here $T = a \sqrt{3 (n^2 + m^2 + nm)} /  \rm{gcd}(2n+m,2m+n)$, is the length of the translation vector (lattice constant) defining the conventional unit cell of the nanotube,   and $N = 2(n^2 + m^2 + nm)/d_R$ denotes 
the total number of A (B) atoms in the conventional 1D nanotube unit cell~\cite{KouwenhovenRMP}.

Having constructed the positions $\br = \br(\bx)= \br(\nu,\ell,\tau)$, as well as the tunneling matrix elements, 
$t(\br,\br')=t(\bx-\bx')$, we can construct and diagonalize the non-interacting part of the Hamiltonian
(3) in the mai text, and obtain the corresponding eigenfunctions 
$\phi_\alpha(\br) \equiv \phi_\alpha(\nu,\ell,\tau)$, and rewrite the noninteracting part of the Hamiltonian as 
\beq
H_{0} = \sum_{\alpha,s}  \epsilon_{\alpha}  c^\dagger_{\alpha s} c_{\alpha s}.
\eeq

Next, we express the interaction in this basis as 
\begin{multline}
  H_{\rm int} = 
  \frac{1}{2} \sum_{\substack{ \\ s_1 s_2}} \sum_{\alpha\beta\gamma\delta  } 
 V_{\alpha\beta ;\gamma\delta } 
  c_{\alpha  s_1}^\dagger
  c_{\gamma  s_2}^\dagger
  c_{\delta s_2}
  c_{\beta  s_1} \\
  +
  \sum_{ s}
  \sum_{\alpha,\beta,\eta}
  \Big (
   V_{\alpha \eta; \eta \beta }
	c_{\alpha  s}^\dagger
	c_{\beta  s}-
   V_{\alpha \beta; \eta \eta }
	c_{\alpha s}^\dagger
	c_{\beta  s}
\Big )	,
  \label{eq:H_U_eigenBasis}
\end{multline}
where the last terms originate from normal ordering, and  $V_{\alpha\beta;\gamma\delta }$
denotes the two-body interaction element, 
\begin{equation}
  V_{\alpha \beta; \gamma\delta}
  = \sum_{a,a'} 
  \phi_{\alpha}^{*}(a)
  \phi_{\beta}(a)  V( \br_a - \br_{a'})
  \phi_{\gamma}^{*}(a')
  \phi_{\delta}(a').
  \label{eq:Vijkl}
\end{equation}
Here, for compactness, we have introduced the composite label, $a=(\nu,\ell,\tau)$.
For  effective Coulomb interaction we use the so-called Ohno potential~\cite{Perebeinos-PRL-2004},
\begin{align}
 V ( \br_1 - \br_2 ) 
  = \frac{e^2}{ \epsilon_r} \frac{1}{ \sqrt{ (\br_1 - \br_2)^2 + {\alpha}^2 } }, \,\nonumber
\end{align}
with ${\alpha} = \frac{e^2}{ \epsilon U_0}$, and $U_0 = 11.3$ eV for 
the $\pi$-orbital~\cite{Perebeinos-PRL-2004}.

To perform density matrix renormalization group (DMRG) 
calculations, we now assume that quantum fluctuations only influence 
the occupation of levels not far from the Fermi energy, and therefore restrict the active space 
of our many-body computations to  states close to the Fermi energy, $|\epsilon_\alpha|<\Lambda$, with $\Lambda $
an energy cut-off introduced. However, in doing so, we must treat occupied 'core' levels and normal ordering 
carefully. 
In practice, we do that by adding   the truncated  normal ordered part of the interactions to the kinetic energy, and thereby renormalizing the non-interacting part as
\be
H_0 \to \tilde H_0 \equiv \sum_s\sum_{|\epsilon_\alpha|,|\epsilon_\beta|<\Lambda } 
T_{\alpha\beta}  c^\dagger_{\alpha s}c_{\beta s}\;,
\ee
with the single particle matrix elements defined as 
\be
T_{\alpha\beta } = \epsilon_\alpha \delta_{\alpha\beta}+ \sum_{|e_\gamma|<\Lambda }\Big (V_{\alpha \gamma;\gamma\beta}-V_{\alpha\beta;\gamma\gamma}\Big), 
\label{eq:Tij}
\ee 
while  interactions are restricted  to active orbitals, 
\be
H_{\rm int} \to \widetilde H_{\rm int} \equiv 
 \frac{1}{2}  \widetilde{\sum_{{\alpha,\beta,\gamma,\delta}  \atop{s_1 s_2} } }
 V_{\alpha\beta ;\gamma\delta } 
  c_{\alpha  s_1}^\dagger
  c_{\gamma  s_2}^\dagger
  c_{\delta s_2}
  c_{\beta  s_1} 
  \;.
\ee
Here the tilde sign indicates restriction to active orbitals. A delicate and  important feature of the cut-off 
construction above  is that it \emph{ preserves } electron-hole symmetry  even for the interacting spectrum,
in case we have only nearest neighbor hopping, as readily verified 
by explicit analytical calculations as well as by our numerics.

\subsection{DMRG calculations with long range Coulomb interactions}\label{sec:DMRG}
\rc{
\begin{figure}[tbh]
 \includegraphics[width=0.8\columnwidth]{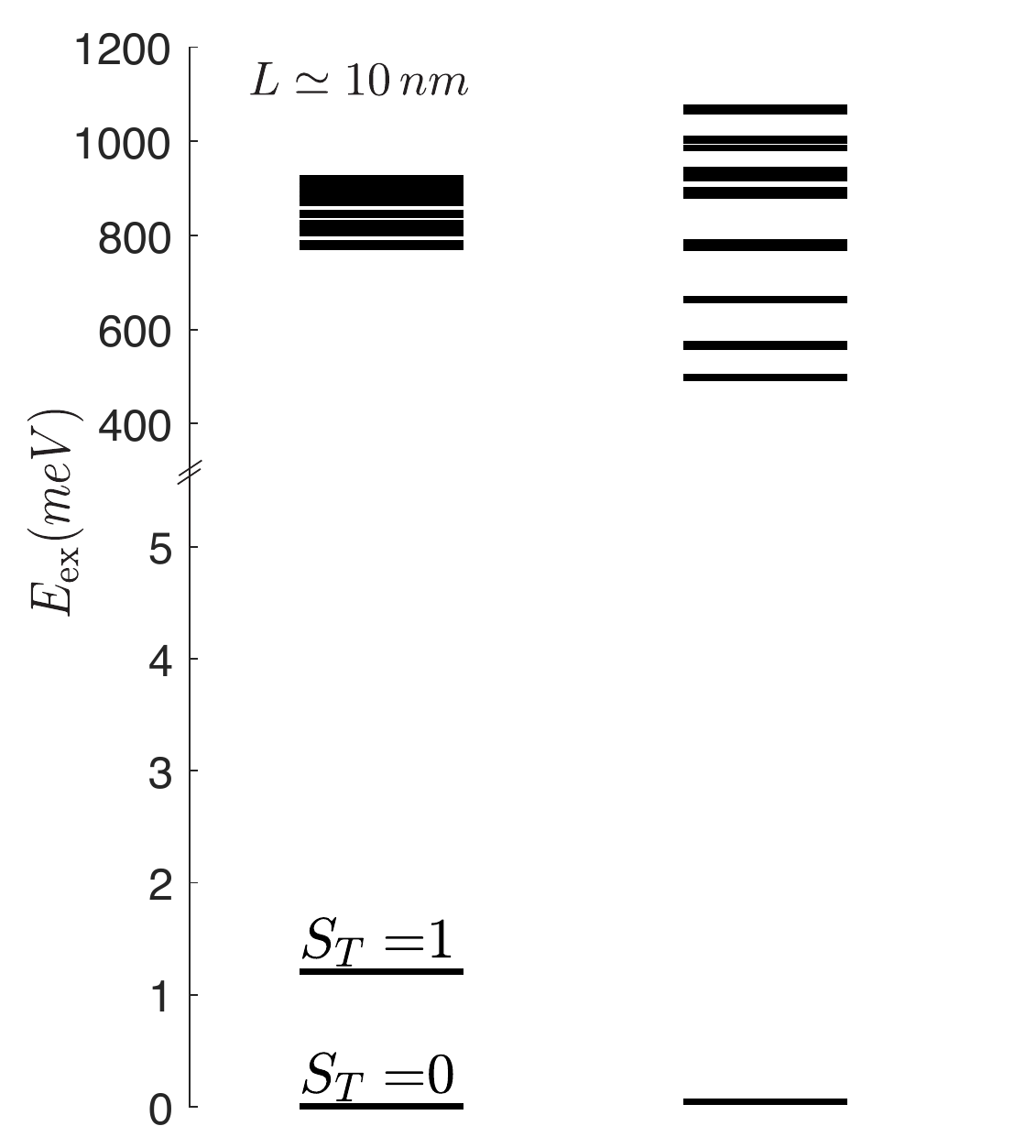}
  \caption{Energy excitation spectrum of a nanotube in
  in the $S_T^{z}=0$ sector,  as obtained with the DMRG. 
 The  chirality of the nanotube is $(n,m)=(7,5)$ and  $L\simeq 10\rm \, nm$. 
 The left columns represent the many body spectrum 
  while the right column represent the non-interacting spectrum.  
  On the left side, the energy separation between the ground state and the first excited 
  state  have been magnified 20 times, for better 
  visibility. The numbers on the two levels are the values of the total spins $S_T$ 
  as extracted from the DMRG calculations.
 }
  \label{fig:energy_levels}
\end{figure}
}
For our DMRG calculations we employ the two-site variant of DMRG, as first introduced by White~\cite{WhiteDMRG}. 
We compute the ground state energy as well as the energy of a few excited states above the ground state.   
In practice, the performance of the DMRG can be boosted significantly by optimizing the computational basis
using fermionic mode transformation\cite{Krumnow}.

The local electron density for the 
effective 1D lattice model  is then easily  expressed in terms of the reduced (spin traced) density matrix 
$\varrho_{\alpha\beta}
\equiv\sum_s  \langle c^\dagger_{\alpha s} c_{\beta s} \rangle$ as 
\be
n(\nu\ell\tau) = \widetilde{ \sum_{\alpha \beta}} \;\varrho_{\alpha\beta} \; \phi_\alpha^ {*}(\nu\ell\tau) \phi_{\beta}(\nu\ell\tau)
+ n^{\rm core}(\nu\ell\tau)\,, 
\nonumber 
\ee
with $n^{\rm core}(\nu\ell\tau)$ the electron charge of the completely occupied core states. 
A summation over the helix label $\nu$ yields the total density of atoms $A$ or $B$ at a helix position $\ell$,
\be
n^\tau(\ell) \equiv \sum_\nu n(\nu\ell\tau)\, .
\ee
Computing then the excess   charge density, 
  $\Delta n^\tau(\ell) \equiv  n_{Q=1}^\tau(\ell) -  n_{Q=0}^\tau(\ell) $, induced upon
 adding one electron to the nanotube,  allows us, for example, to explore the 
localization and spin structure of the edge states and 
to estimate their extension in real space, as presented in Fig.~4 of the main text.
\rc{In Fig.~\ref{fig:energy_levels} we represent the lowest part of the energy spectrum 
for a nanotube with chirality $(n,m)=(7,5)$. It allows us to extract the exchange interaction 
between the localized spins at the two ends of the nanotube. 
For  (7,5) chirality,  the ground state is 
antiferromagnetic with a total spin $S_T=0$.  
In general, two spins of size $S$, coupled by the Hamiltonian (2) , has an spectrum $E_n =J_{\rm eff}
(S_T(S_T+1)-2S(S+1))/2$. The energy difference between the first excited state ($S_T=1$) and the ground 
state ($S_T=1$) is therefore $\Delta E =  J_{\rm eff}$, which allows us to extract $J_{\rm eff}$ 
directly from the DMRG spectrum. In case of larger end spins, further low-lying excitations appear with energies, 
which we can also clearly see in the spectrum.
}


\end{document}